\documentclass[prd,a4paper,11pt,nofootinbib,showkeys,showpacs]{revtex4}

\usepackage[frenchb,english]{babel}
\usepackage[applemac]{inputenc}
\usepackage{enumerate}
\usepackage{amsmath}
\usepackage{amsthm}
\usepackage{amssymb}
\usepackage{graphicx}
\usepackage{floatflt}
\usepackage{fancybox}
\usepackage{stmaryrd}
\usepackage{appendix} 

\usepackage{hyperref} 

\usepackage{color}

\renewcommand{\Re}{\textrm{Re}}

\newcommand{\Ai}{\textrm{Ai}}

\newcommand{\om}{\omega}
\newcommand{\Om}{\Omega}
\newcommand{\bom}{\bar \omega}

\newcommand{\as}{{\rm as}}

\newcommand{\mev}[1]{\langle #1 \rangle}
\newcommand{\vect}[1]{\overrightarrow{#1}}
\newcommand{\p}{\partial}

\newcommand{\grad}{\textrm{\bf grad}}

\newcommand{\be} {\begin{equation}}
\newcommand{\ee} {\end{equation}}
\newcommand{\bsub}{\begin{subequations}}
\newcommand{\esub}{\end{subequations}}
\newcommand{\bea}{\begin{eqnarray}}
\newcommand{\eea}{\end{eqnarray}}
\newcommand{\bi} {\begin{itemize}}
\newcommand{\ei} {\end{itemize}}
\newcommand{\ben} {\begin{enumerate}}
\newcommand{\een} {\end{enumerate}}
\newcommand{\bmat} {\begin{pmatrix}}
\newcommand{\emat} {\end{pmatrix}} 
\newcommand{\bal} {\begin{aligned}}
\newcommand{\eal} {\end{aligned}}
\newcommand{\btab}{\begin{tabular}}
\newcommand{\etab}{\end{tabular}}

\newcommand{\eq}[1]{Eq.~\eqref{#1}}

\begin{document}
\selectlanguage{english}

\title{Undulations from amplified low frequency surface waves}

\author{Antonin Coutant}
\email{antonin.coutant@aei.mpg.de}
\affiliation{Max Planck Institute for Gravitational Physics, Albert Einstein Institute, Am Muhlenberg 1, 14476 Golm, Germany, EU}

\author{Renaud Parentani}
\email{renaud.parentani@th.u-psud.fr}
\affiliation{Laboratoire de Physique Th\'eorique, CNRS UMR 8627, B\^atiment 210, Universit\'e Paris-Sud 11, 91405 Orsay Cedex, France}

\date{\today}

\begin{abstract}
We study the linear 
scattering of gravity waves in longitudinal inhomogeneous stationary flows. 
When the flow becomes supercritical, it is known that counterflow propagating shallow 
waves are blocked and converted into deep waves. 
Here we show that in the zero-frequency limit,
the reflected waves are amplified in such a way that the free surface develops an undulation, 
i.e., a zero-frequency wave of large amplitude with nodes located at specific places. 
This amplification involves negative energy waves, and 
implies that flat surfaces are 
unstable against incoming perturbations of arbitrary small amplitude. 
The relation between this instability and black hole radiation 
(the Hawking effect) is established. 
\end{abstract}

\maketitle

\newpage
\section{Introduction}

It has long been observed that stationary flows that 
become supercritical, i.e., when the flow velocity equals the speed of low frequency
surface waves, are often associated with an undulation, 
i.e., a zero-frequency wave with a macroscopic amplitude~\cite{Johnson,Lawrence87}. 
Undulations have been extensively studied, often 
in the forced and in the nonlinear regime, see e.g.~\cite{Wu87,LeePhD,Byatt71,El05,El05b}. 
In this paper, we study their appearance in another regime, namely from the scattering of low frequency shallow waves which propagate against 
a background flow with a flat surface, i.e. when the forcing vanishes. In this case, we shall show that 
an undulation develops because the background flow is unstable against incoming waves with arbitrary small amplitudes. 
In this we have been inspired by the fact that 
their scattering near the blocking point is akin to that governing the Hawking effect~\cite{Hawking75,Unruh76,Unruh81,Unruh95,Schutzhold02,Balbinot06}, 
which predicts that black holes should spontaneously emit a thermal flux. 

To understand the scattering of counterflow waves, one must take into account their 
dispersion relation. In homogeneous stationary flows, when neglecting capillary effects~\cite{Johnson}, 
the relation between $\Om$ (the frequency measured in the fluid frame) and the wave vector $k$  is given by   
\be
\Om^2 = g k\tanh(h_B k) , 
\label{Fdr}
\ee
where $h_B$ is the height of the background flow, and $g$ the gravitational acceleration. 
When the flow is inhomogeneous, at fixed frequency $\om$ measured in the lab frame, 
the co-moving frequency $\Om$ becomes a function of $x$ given by 
$\om = \Om + v_x(x)k_\om$, where $v_x(x)$ is the longitudinal velocity flow,  
and $k_\om$ the $x$-dependent wave vector. 
When $\om$ is high enough, a counter flow wave packet is blocked and generates 
two reflected wave packets, a long wavelength co-propagating mode, and a short wavelength one which is dragged by the flow. 
In this case, the (positive) incoming energy is shared among the two outgoing waves. 
When lowering $\om$ below of certain critical frequency $\om_\kappa$, which is related to the gradient of $v_x(x)$ 
evaluated when $F_n$ crosses 1, a third wave 
acquires a non-negligible amplitude. (We define
the Froude number $F_n$ as the ratio of the flow velocity $v_x$ over the speed of 
low frequency waves.) 
This extra wave possesses a negative energy, which means that 
we are now facing an {\it over-reflection}~\cite{Acheson76}, i.e., an amplification process.\footnote{
Such scattering is also said {\it anomalous} because the energy carried by the other  
outgoing waves is {\it higher} than the incoming wave energy. 
Negative energy waves are also known in particle physics and  
quantum field theory. In these contexts, when mixing with positive energy waves, 
they are responsible for spontaneous pair creation effects~\cite{Greiner,BirrellDavies,Primer}.
While negative energy waves can often be ignored (because they do not significantly mix
with the positive energy waves), they are responsible for various types of instabilities, 
see~\cite{Fabrikant} for examples in shear flows.} 
In a recent experiment~\cite{Weinfurtner10,Weinfurtner13}, 
the production of this extra wave have been clearly observed 
in the linear regime we shall use. 

In the present paper, we study a limit which was not studied in~\cite{Unruh81,Schutzhold02,Rousseaux07,Weinfurtner10,Unruh12}. 
It concerns the limit $\om \ll \om_\kappa$. 
In this case, the long wavelength co-propagating mode 
plays no significant role, while 
unusual properties characterize the two short wavelength modes of opposite energy. 
First, in the limit $\om \to 0$, they acquire the {\it same} amplitude,  
merge, and form a single standing wave with zero frequency and nodes at definite places. 
Second, the amplification factor diverges as $1/\om$. 
Using the correspondence with black hole geometry, we show that 
this divergence is directly related to the famous prediction of Hawking radiation~\cite{Hawking75}. 
Using the fact that perturbations of inhomogeneous flows propagate as light waves on a curved space-time~\cite{Unruh81},
one realizes that the supercritical flows we consider correspond to ``acoustic white holes''~\cite{Schutzhold02,Rousseaux07,Weinfurtner10,Unruh12} with their  
horizon located where $F_n=1$. 
In fact, the generation of undulations 
and black hole emission are based on a common amplification mechanism. 
This explains why undulations have been observed 
in the experiments~\cite{Rousseaux07,Weinfurtner10,Weinfurtner13} 
aiming at measuring the analogue Hawking effect. 

The paper is organized as follows. 
In Section II, we present the wave equation, discuss the scattering 
of stationary waves in supercritical flows, and demonstrate that the amplification factor diverges as $1/\omega$ for $\omega \to 0$. 
In Section III, we study incoming waves packets and 
show that the two reflected waves merge and form a single undulation 
in the limit $\om \to 0$. 
We then show that incident low frequency waves with random properties also give rise to the same undulation, but with a growing amplitude. In Appendix \ref{UnruhApp}, we derive the wave equation and relate it to the relativistic equation used by Hawking.
In Appendix \ref{energyApp}, we review the main properties of 
the conserved inner product which governs the amplification process. 
In Appendix \ref{HRApp}, we explain how to compute the amplification factor without having recourse to standard WKB techniques
which fail in the present case. 

\section{Settings}

In this Section, we adapt to the present case results which have been recently obtained in 
other works. The important new results are presented in the next Section.  

\subsection{Wave equation et action formalism}

We consider surface waves which propagate in a water tank of constant transverse dimension $L_\perp$. 
We assume that the flow is incompressible, non turbulent, and irrotational. 
We also assume that both the bottom of the tank and the background free surface 
do not depend on the transverse coordinate, and become asymptotically flat in the upstream region.  
We call $h_\as$ and $v_\as$ the asymptotic values of the water depth and the background
flow velocity in this region. For simplicity, we only study waves with no dependence in the transverse coordinate\footnote{Modes with a non zero transverse momentum $p_\perp$ are studied in~\cite{Coutant12}.}, and we neglect the effects of capillarity. To incorporate the latter, one should consider the dispersion relation which generalizes \eq{Fdr} by including the capillary length~\cite{Schutzhold02}.
Including these short wavelength effects will not affect the main conclusions of our work. 

When the background flow is non-uniform, the linear equation for surface waves 
is rather complicated~\cite{Schutzhold02,Unruh12}. 
The origin of the difficulty stems from the fact that we aim to study the zero frequency limit. As a result, we cannot use 
the standard slowly varying (WKB) approximation where 
the wave vector $k_\om $ is much larger than the typical spatial gradient of the background flow. 
In Appendix~\ref{UnruhApp}, we recall the main steps to obtain it and compare it with other models of water waves. In the body of the text, we shall exploit the fact that 
this equation can be derived from an action. 
Interestingly, this action possesses a rather simple structure which, moreover, 
is very similar to that describing sound waves in an irrotational fluid~\cite{Balbinot06}.\footnote{It should be emphasized that 
undulations occur in other dispersive media. In fact they 
are closely related to the ``layered structures''  found in$~^4$He~\cite{Pitaevskii84}, and in Bose gases~\cite{Baym12}, when the flow exceeds the Landau critical velocity. 
They also occur in supersonic flows 
 liquid helium~\cite{Rolley07}, and in atomic Bose Einstein condensates~\cite{Mayoral11,Coutant11} 
where the dispersion relation is $F_{\rm BEC}^2(k) = k^2 + \xi^2 k^4/4$ instead of \eq{Fdr}.  
In the present paper, even though we have in mind gravity waves, we formulate the problem in terms which apply to the general case. } 
In addition, the action formalism is appropriate to efficiently describe the wave scattering, as well
as to establish the relationship with the Hawking treatment 
of black hole radiation. 

The action for the perturbations $\phi$ of the velocity potential has the following structure
\be
S= \frac{1}{2}\int \rho(x) \left\{ \frac1{c^2(x)}  [(\partial_t + v_x(x) \, \partial_x)\phi]^2 - \phi \, \hat F^2(d_\Lambda(x)) \, \phi \right\} dx dt. 
\label{Sact}
\ee
The function $\rho(x)$ is an effective 1-dimensional fluid density, $v_x(x)$ is the background flow velocity, 
$c(x)$ fixes the low frequency group velocity, and $\hat F^2$ 
is a differential operator which governs the dispersion relation.
The combination $\rho \hat F^2$ forms a self-adjoint operator, and 
$d_\Lambda(x)$ is the local dispersive wavelength. (In an atomic Bose condensate, the latter is 
known as the healing length~\cite{Dalfovo99,Macher09b}.) 
For each fluid, these functions are related in a specific manner to the properties of the background flow. 

For gravity waves, assuming an incompressible fluid, i.e., a constant 3-dimensional density $\rho_0^{3D}$, 
there are several (physically equivalent) ways to identify these 
functions. Using the results of Appendix~\ref{UnruhApp}, 
a convenient choice is
\bsub \bea 
\hat F^2(d_\Lambda(x)) &=& \frac{1}{d_\Lambda(x)} i\partial_x \tanh (d_\Lambda(x) i\partial_x ),
\label{Fd} \\
d_\Lambda(x) &=& h_\as \frac{v_\as v_x(x)}{v^2(x)} , \label{dLambda}\\
\rho(x) &=& \rho_0^{3D} L_\perp d_\Lambda(x), 
\\  c^2(x) &=& d_\Lambda(x) \left[ g + \frac{v_x^2}{v^2} \partial_y (v^2/2)_{y\, =\, h_B(x) } \right] \doteq d_\Lambda(x) \, g_{\rm eff}(x). 
\label{c_expr}\eea 
\label{gwexpressions}
\esub
In the above, $v^2 = v_x^2 + v_y^2$, where $v_x$ and $v_y$ are the horizontal and vertical components of the background velocity, 
evaluated along the free surface $y= h_B(x)$. 
(Note that the local Froude number is now unambiguously defined as $F_n = v_x(x) / c(x)$.) 
The quantity $g_{\rm eff}$ is the effective gravitational acceleration 
which takes into account the centrifugal acceleration.  
Asymptotically, all $x$ dependence are negligible, and \eq{c_expr} delivers the standard expression $c_\as^2 =h_\as g $. 
For long wavelengths, i.e. low gradients $k d_\Lambda \ll 1$, the dispersive length $d_\Lambda$ drops out from \eq{Fd} 
and one gets the dispersionless expression $F^2 \to k^2$, since $k = -i\partial_x$. 
For smaller wavelengths, combining $F^2$ and $c^2$, 
one finds a generalized version of \eq{Fdr} 
where $d_\Lambda$ acts as a dressed value of $h_\as$. 
In fact, to lowest order in $\p_y v_x$, one can show that $d_\Lambda(x)$ reduces to the water depth at $x$ 
(see \eq{WKBdepth} in Appendix~\ref{UnruhApp}). 

As far as the scattering of waves is concerned, all we need 
is the knowledge of the differential operator $\hat F^2$ and the 
functions  $\rho,v,c,d_\Lambda$ entering in \eq{Sact}. 
In other words, the intricate aspects of the above equations will play no significant 
role in the sequel.
Yet, to make physical predictions, and to test them, 
one needs the relation between the velocity potential $\phi$ and 
the vertical fluctuation of the surface $\delta h(t,x) = h(t,x) - h_B(x)$
with respect to the background free surface $y = h_B(x)$. 
As explained in Appendix \ref{UnruhApp}, see \eqref{Appdeltah}, 
this relation is 
\be
\delta h(t,x) = - \frac{1}{g_{\rm eff}(x)} (\p_t + v_x(x) \p_x) \phi(t,x) .
\label{deltah} 
\ee
It is interesting to notice that $\delta h(t,x)$ is 
related by a constant factor to $\pi(t,x)$, the momentum conjugated to $\phi(t,x)$ given the action of 
\eq{Sact}~\footnote{Notice that in the standard  Hamiltonian formulation of surface wave propagation, 
the variable is $\delta h$, and $\phi$ its conjugate momentum~\cite{Dingemans}. Here we have chosen to work in the opposite 
convention as the relation to the relativistic space-time description is  
straightforward using $\phi$, see App.\ref{UnruhApp}.}. 
Indeed, taking the variation of the action with respect to $\p_t\phi$, one obtains
\bsub \bea
\pi(t,x) &\doteq& \frac{\rho(x)}{c^2(x)} (\p_t + v_x(x) \p_x) \phi(t,x) , \label{H1}\\
&=& -(\rho_0^{3D} L_\perp)\,  \delta h(t,x) . \label{H2}
\eea \esub
It is also interesting to notice that for sound waves, e.g. in an atomic Bose condensate, the density fluctuation $\delta \rho$ is related to the momentum $\pi$, and the potential $\phi$, by very similar 
equations, see Eq.~(B.11) in~\cite{Macher09b}. 
Therefore the forthcoming analysis also applies to these waves, when using the appropriate $\hat F^2$ operator governing dispersion. 

\begin{figure}[!ht]
\begin{center}
\includegraphics[width=\columnwidth]{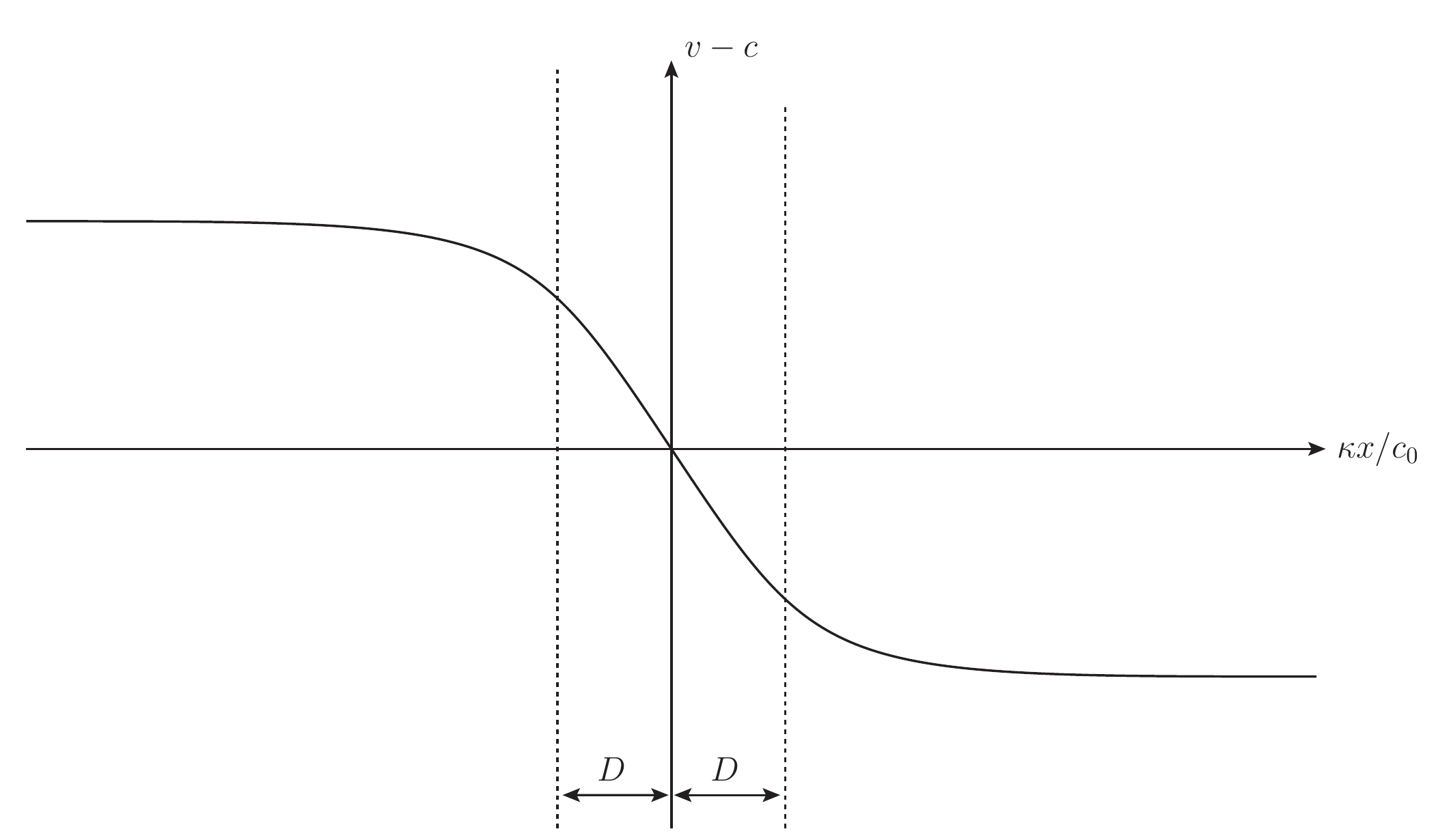}
\end{center}
\caption{Background profile $v_x-c$ as a function of $x$ with $v > 0$. 
The chosen profile is $v_x - c = - c_0 D \tanh\left({\kappa x}/{c_0 D} \right)$, 
where $c_0$ is the speed at $x= 0$. 
The parameter $D$ governs the extension of the near wave blocking region 
where $c-v_x \sim \kappa x$. 
The frequency $\kappa$ is defined by $\p_x (v_x - c)_{|x=0} = -\kappa$. 
As explained in App.~\ref{HRApp} it governs both the spatial properties of the modes in this region, 
and the non-trivial mode mixing. In a gravitational context, it is known as 
the surface gravity~\cite{Unruh81}. On the asymptotic subcritical right side, $v_x(x)$ and $c(x)$ reach the 
constant values $v_\as $ and $c_\as$. }
\label{vprofile_fig}
\end{figure}

When applying the Legendre transform ($H \doteq \pi \partial_t \phi - \mathcal{L}$) 
to the action density of \eq{Sact}, one obtains the Hamiltonian 
\be
H = \frac12 \int \frac{\rho}{c^2} \left\{ \left[\frac{c^2 \pi}{\rho} -  v_x \, \p_x\phi\right]^2 - v_x^2 (\p_x \phi)^2 + c^2\,  \phi \hat F^2(d_\Lambda) \phi \right\} dx .
\label{hamilt}
\ee
For stationary flows, $H$ is conserved and furnishes the energy carried by the waves.
For homogeneous backgrounds, \eq{hamilt} coincides with standard expression~\cite{Dingemans}. 
The wave equation can then be obtained from Hamilton equations. The first equation 
$\partial_t\phi = \{\phi, H\}$, where $\{ \, , \}$ is the Poisson bracket, gives back 
\eq{H1}. The second equation, $\partial_t\pi = \{\pi, H\}$, gives
\be
(\p_t + \p_x v_x(x) )\pi(t,x) = - \rho(x) \hat F^2(d_\Lambda(x)) \, \phi(t,x), \label{H2} 
\ee
which corresponds to Eq.~(86) in~\cite{Unruh12}. Taken together, these equations give
\be
\left[ (\p_t + \p_x v_x(x) )\frac{\rho(x)}{c^2(x)} (\p_t + v_x(x) \p_x) + \rho(x) \hat F^2(d_\Lambda(x)) \right] \phi(t,x) = 0.
\label{twaveeq}
\ee
The forthcoming analysis is based on this wave equation
applied to supercritical flows as that depicted on Fig.\ref{vprofile_fig}.
We emphasize that we shall neither make use 
of the standard slowly varying approximation, i.e.,  $\kappa/c_0 \ll k_\om$
in term of the typical frequency defined in the Figure caption,
nor assume that the flow is near critical, (i.e., $|F_n - 1| \ll 1$). 
Our treatment thus applies to {\it arbitrary low} frequencies, and is valid in flows where $v$ and $c$ significantly vary. 
This contrasts with the regimes commonly explored, e.g. in~\cite{Johnson,Mei,Dingemans} (for further discussion on this, see App.~\ref{valid_App}). 

\subsection{Incoming and outgoing mode bases} 
\label{NEW_Sec}

To study the scattering of monochromatic waves of fixed frequency $\om$, we 
introduce the {complex} modes $e^{- i\om t}\phi_\om(x)$. 
\eq{twaveeq} implies that their spatial part obeys 
\be
\left[(\om + i\p_x v_x(x) )\frac{\rho(x)}{c^2(x)} (\om + i v_x(x) \p_x) - \rho(x) \hat F^2(d_\Lambda(x)) \right] \phi_\om(x) = 0 . 
\label{omwaveeq}
\ee
The real time-dependent wave is then given by 
$\phi(t,x) = 2\Re\left(e^{- i\om t}\phi_\om(x)\right)$. To describe the scattering, 
we use two different mode bases, for more details in a similar context see~\cite{Macher09}. 
The {\it in} basis describes modes that shall be scattered, while the {\it out} one describes modes that have been scattered. 
These modes are identified through the standard procedure~\cite{Greiner,Fulling,Primer}: in the past (resp. in the future),
each incoming mode $\phi_\om^{\rm in}$ (resp. outgoing $\phi_\om^{\rm out}$) asymptotes in the sense of a broad wave packet 
to a single plane wave with a group velocity $v_g = (\p_\om k)^{-1}$ directed toward (resp. away from) the blocking point. 
These asymptotic plane waves are given by $\sim e^{i k_\om^a x}$ where $k^a_\om$ is a 
real root of the dispersion relation 
\be
(\om - v_\as k^a_\om )^2 = g k^a_\om \tanh\left(h_\as k^a_\om \right)  = (\Om^a_\om)^2 . 
\label{HJ}
\ee
As a result, the number of independent modes is equal to the number 
of real solutions of \eq{HJ} with $\om > 0$. 

\begin{figure}[!ht]
\begin{center}
\includegraphics[width=\columnwidth]{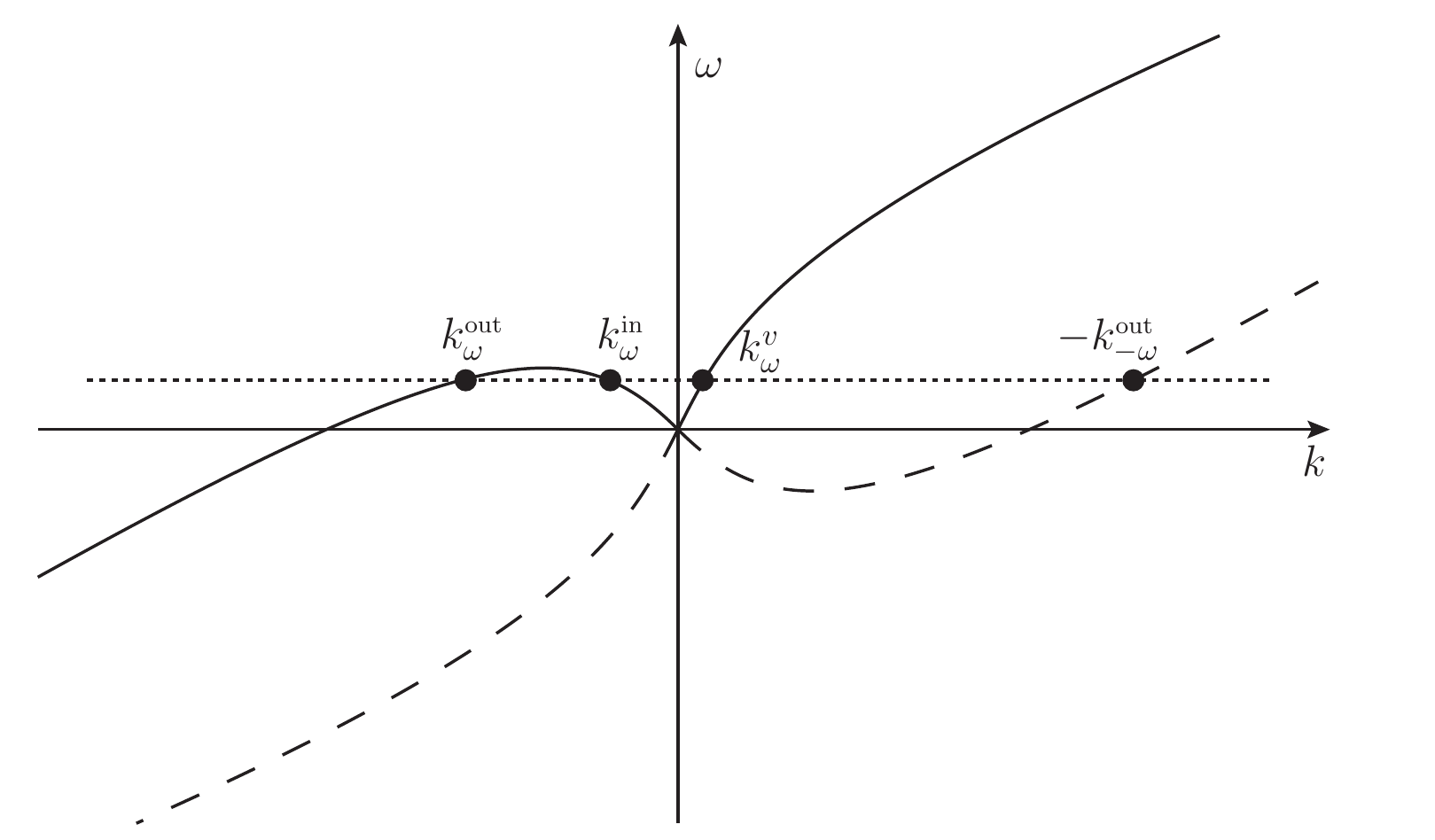} 
\end{center}
\caption{Dispersion relation of \eq{HJ} in the $\om - k$ plane when $v_\as > 0 $ and sub-critical $v_\as < c_\as$. There are two double branches: 
that with $\Omega > 0$ (continuous lines) characterizes the {\it positive norm modes}, see App.~\ref{energyApp}, 
while the dashed lines describe the {\it negative norm modes}. 
For a low frequency $\om > 0$, there are four roots. One sees that only 
$-k_{-\om}^{\rm out} > 0$ lives on the second branch. As explained in the text, 
the corresponding mode $(\phi_{-\om}^{\rm out})^*$ carries negative energy, and its 
mixing with the other modes leads to an over-reflection. One also sees that in the limit $\om \to 0$, $-k_{-\om}^{\rm out} $ and $k_{\om}^{\rm out} $
reach $\pm k_Z$ respectively. As shall be shown, $k_Z$ is the wave vector of the undulation. 
}
\label{Disprel_fig}
\end{figure}

For this reason, it is worth studying these roots. 
On the right asymptotic side, where the flow is subcritical ($|v_\as| < c_\as$), there is a threshold value $\om_{\rm max}$ 
which separates two cases. 
For $\om$ above $\om_{\rm max}$, there are 2 real roots, as in the absence of a flow. 
Instead, in the low frequency regime which interests us, for $0 < \om < \om_{\rm max}$, there are 4 real roots, see Fig.~\ref{Disprel_fig}, which means that there two new types of stationary waves. 
Because each asymptotic root corresponds to either an {\it in} or {\it out} mode,
we call the various roots using the name of the corresponding mode.
For instance, $k_\om^{\rm in}$ describes the (usual) 
long wavelength incoming left moving mode $\phi_\om^{\rm in}$.
When $\om \to 0^+$, one has 
\be
k_\om^{\rm in} \sim -\frac{\om}{c_\as - v_\as} = \om/v_g^{\rm in} < 0. 
\ee
The group velocity $v_g^{\rm in} = - c_\as + v_\as < 0$ confirms that it is moving leftward. 
The positive long wavelength root $k_\om^{\rm co, \, out}$ describes the  (usual) co-moving 
outgoing mode $\phi_\om^{\rm co,\, out}$  since $v_g^{{\rm co,\, out}} = c_\as + v_\as > 0$. 
The last roots $k_\om^{\rm out} < 0 $ and $-k_{-\om}^{\rm out} > 0$ are the two new ones. They both 
correspond to short wavelengths conter-propagating modes which are both 
swept along with the flow. When $\om \to 0$, as clearly seen in Fig.~\ref{Disprel_fig}, 
they reach opposite value $\mp k_Z$ respectively, with $k_Z > 0$. 
More precisely, to first order in $\om$  
\be
k_\om^{\rm out} = - k_Z + \om/v_g^Z, \label{kZ}
\ee
where $v_g^Z>0$. 

In Fig.~\ref{Disprel_fig}, one also notices that three of the four roots, namely 
$k_\om^{\rm in}$, $k_\om^{\rm co,\, out}$ and $k_\om^{\rm out}$, live on the  
branch of solutions with a positive comoving frequency $\Omega_\om = \om - v_\as k_\om$. 
This branch characterizes the modes with positive energy and {\it positive norm}, 
see App.\ref{energyApp} for more details concerning this important aspect. 
Instead, the fourth root has a negative energy, and lives in the ``unusual'' branch of solutions of negative $\Omega$
and {\it negative norm}. To keep this in mind, we call this root $-k_{-\om}^{\rm out}$. 
First, because $k_{-\om}^{\rm out} < 0$ lives on the usual branch
when considering the opposite value of $\om$, and second, 
because the curves of Fig.~\ref{Disprel_fig} are left invariant under both $\om \to - \om$ and $k \to -k$, which replaces the positive $\Omega$ branch by the negative one, and vice versa.
The outgoing modes which correspond to $k_{\om}^{\rm out}$ 
and $-k_{-\om}^{\rm out}$ shall be respectively called $\phi_\om^{\rm out}$ for the positive norm one,
 and $\left(\phi_{-\om}^{\rm out}\right)^*$ for the negative norm one. 
 
In usual circumstances, i.e. when $\om$ is not too small, the mixing of 
$\left(\phi_{-\om}^{\rm out}\right)^*$ with the positive norm modes is so small  
that it can be safely ignored. In this case, one deals with an elastic scattering. 
However, at low frequencies, for supercritical flows like that of Fig.~\ref{vprofile_fig}, 
the mixing becomes so important that it is must be included 
to account for the observed phenomena~\cite{Rousseaux07,Weinfurtner10,Weinfurtner13}. 
On the other hand, the co-propagating mode 
$\phi_\om^{\rm co}$ plays essentially no role~\footnote{Notice that neglecting this mode is also 
done in weakly nonlinear treatments when passing from 
the Boussinesq to the Korteweg-de Vries equation.}. 
Hence, we are effectively facing an ``over-reflection'' involving 
only two pairs of modes. 

The two {\it out} modes $\phi_\om^{\rm out}, (\phi_{-\om}^{\rm out})^*$, and the {\it in} mode $\phi_\om^{\rm in}$ 
have been already described. The last one is the negative norm mode $(\phi_{-\om}^{\rm in})^*$. At early times, it 
asymptotically describes an incoming mode which comes from the left super-critical region. 
The corresponding root is negative and called $-k_{-\om}^{\rm in}$ because it has a negative $\Omega$,
and carries a negative energy. 

\subsection{Low frequency mode amplification} 
\label{scat_Sec}

Since the {\it in} modes, and the {\it out} modes, 
form two basis of solutions of \eq{omwaveeq} (when neglecting the co-propagating mode), they are related by a linear transformation, 
\be
\bmat \phi_\om^{\rm in} \\ (\phi_{-\om}^{\rm in})^* \emat = \bmat \alpha_\om & \beta_\om \\ \tilde \beta_\om & \tilde \alpha_\om \emat \cdot \bmat \phi_\om^{\rm out} \\ (\phi_{-\om}^{\rm out})^* \emat. \label{Smat}
\ee
Because the modes have opposite norm, the coefficients obey the anomalous scattering relation 
\be
|\alpha_\om|^2 - |\beta_\om|^2 = |\tilde \alpha_\om|^2 - |\tilde \beta_\om|^2 = 1, \label{U11rel}
\ee
in the place of the standard relation $|R_\om|^2 + |T_\om|^2= 1$ between the reflection
and transmission coefficients $R_\om$ and $T_\om$. 
\eq{U11rel} implies that $|\alpha_\om|>1$, which means that the scattering 
leads to an amplification of the waves.

The real task is to compute the coefficients $\alpha_\om$ and $ \beta_\om$. 
It turns out that for low frequencies a proper evaluation in background profiles as in Fig.~\ref{vprofile_fig} is non trivial. 
Indeed, the standard WKB treatment gives 
$\beta_\om \equiv 0$, which means that modes of opposite norm 
(and energy) do not mix. To obtain $\beta_\om $ 
one should therefore use more involved technics. These are presented in 
Appendix \ref{HRApp}. The main result is as follows.
When the supercritical flow is smooth enough, 
\be
\frac{|\beta_\om|^2}{|\alpha_\om|^2} = e^{-\frac{2\pi \om}\kappa},
\label{HRbeta}
\ee
where $\kappa$ is the background flow frequency defined in the caption of Fig.~\ref{vprofile_fig}. It is worth mentioning that 
\eq{HRbeta} was found by Hawking~\cite{Hawking75} in a gravitational context 
when he established that incipient black holes should emit a thermal spectrum at a temperature
given by $k_BT_H = \hbar \kappa/2\pi$. A careful comparison~\cite{Unruh95,Brout95,Coutant11} 
confirms the close connection
between the scattering of waves on a supercritical flow and in a black hole geometry.
The main difficulty is to properly include dispersive effects. 
In Appendix \ref{HRApp}, following~\cite{Coutant11}, we 
recall that, to leading order in $\kappa d_\Lambda/c_0 \ll 1$, 
the replacement of the dispersion relation
from the relativistic one used by Hawking 
to that of \eq{Fdr} does {\it not} affect \eq{HRbeta}. 

For frequencies larger than $\kappa$, \eq{HRbeta} 
implies that the negative energy mode has 
an amplitude exponentially reduced with respect to that of the standard wave, something which
has been recently verified~\cite{Weinfurtner10,Weinfurtner13}. This is the standard 
adiabatic regime where modes of opposite norms do not significantly mix. 
Instead, in the opposite limit $\om \to 0$, Eqs.~\eqref{U11rel}, \eqref{HRbeta} 
imply that the coefficients diverge as  
\be
|\alpha_{\om}|^2 \sim |\beta_{\om}|^2 \sim \frac{\kappa}{2\pi \om} \underset{\om \ll \kappa}{\gg} 1. \label{Bogodiv}
\ee
We here underline that this divergence 
is still found when the mixing with the co-propagating 
mode is not negligible and taken into account, 
and also when the inequality  $\kappa d_\Lambda/c_0 \ll 1$ used in Appendix \ref{HRApp}
is no longer satisfied. 
In these cases, $\kappa$ is replaced by a frequency  
$\bar \kappa$ which depends on several quantities, 
see the numerical and analytical work of~\cite{Finazzi12}. 
To indicate that our forthcoming analysis covers these cases as well, we shall use the symbol $\bar \kappa$.

\section{Scattering of low frequency wave packets}

We combine the various elements of the former Section to show that flat free surfaces
are unstable against incoming low frequency waves of arbitrary small amplitude.
As a result, an undulation develops.

\label{wp_Sec}
\subsection{Incoming waves} 

Using stationary {\it in} modes, the general time-dependent solution of \eq{twaveeq} can be written as
\be
\bar \phi(t,x) = 2\Re \left\{\int_0^{\om_{\rm max}} e^{-i \om t} \left[ a_\om \,\phi_\om^{\rm in}(x)  + b_\om \, (\phi_{-\om}^{\rm in}(x))^* \right] d\om\right \},
\label{awp}
\ee
where the coefficients $a_\om$ and $b_\om$ weigh the contribution of
$\phi_\om^{\rm in}$ and $(\phi_{-\om}^{\rm in}(x))^*$.
We now consider a series of incoming wave packets of positive energy, 
sent from the right side against the flow. 
This means that $b_\om = 0$. 
To get explicit expressions, we work with 
\be
a_\om = \frac{A}{n_{\bom}^{1/2}} \exp\left({- \frac{(\om - \bom)^2}{2\sigma_0^2 \bom^2}}\right) , 
\label{aom}
\ee
where $A$ is a complex dimensionless amplitude $A = |A|e^{i \delta}$, 
and where the packets are normalized by  
\be
\int_0^{\om_{\rm max}} |a_\om|^2 d\om = |A|^2.
\ee
We also assume that the waves are almost monochromatic. Irrespectively of the value of $\bom$, this is realized if
\be
\sigma_0 \ll 1. 
\label{packetregime}
\ee 
In this regime, one finds $n_{\bom} = \sigma_0 \bom \pi^{1/2}$. 
The real character of \eq{aom} guarantees that the packets are  
centered around $x=0$ at $t=0$. 

Using the asymptotic behavior of $\phi_\om^{\rm in}$ given in \eq{plw} we get 
\be 
\bar \phi(x,t\to -\infty) = 2\Re \left\{ \bar N A \int_0^\infty e^{- \frac{(\om - \bom)^2}{2\sigma_0^2 \bom^2}} \frac{e^{-i (\om t- k_\om^{\rm in}x)}}{\sqrt{4\pi 
\om c_\as }} \frac{d\om}{n_\sigma^{1/2} } \right\} . \label{Rwp}
\ee
In the broad wave regime of \eq{packetregime}, we can accurately
evaluate the integral with a saddle point approximation. Then, using \eq{deltah} asymptotically, where $g_{\rm eff} = g$, the corresponding incoming 
height variation is
\be
\delta \bar h^{\rm in}(x) \sim \delta \bar
h^{\rm in}_A \times \sin\left(\bom t- k_{\bom}^{\rm in} x + \delta \right) e^{- \frac{\sigma_0^2 \bom^2}2 (t-x/v_g^{\rm in})^2}, \label{indeltah}
\ee
where the amplitude is
\be
 \delta\bar h^{\rm in}_A = |A| \bar N \frac{\bom h_\as}{c_\as - v_\as} \sqrt{\frac{2 \sigma_0}{\pi^{1/2} c_\as^3}} .\label{indeltahAmpl}
 \ee
As expected, the incoming wave \eqref{indeltah} 
oscillates at a frequency $\bom$, has a wavenumber $k_{\bom}^{\rm in}$, 
and its envelope is propagating toward the wave blocking region at a speed $v_g^{\rm in} < 0$. 
We see that the argument of $A$, the phase $\delta$, governs the precise initial positions of the nodes. 
We also see that at fixed $A$, its amplitude $ \delta \bar h^{\rm in}$
decreases linearly for with $\bom$. In this respect, it
is also instructive to evaluate the conserved energy transported by the wave packet. 
Using \eq{hamilt} with \eq{Rwp}, one finds 
\be
\bar E = 2\bom  N |A|^2. 
\label{Enew}
\ee
At fixed $A$, the wave energy linearly vanishes in the limit $\bom \to 0$. 

\subsection{Outgoing waves}

At a time near $t=0$, the packet of \eq{Rwp} reaches 
the wave blocking region around $x=0$, where it undergoes a nontrivial scattering which is governed by \eq{Smat}.
Then two outgoing wave packets are generated and propagate to the right, see Fig~\ref{Contour_fig}.
To analyze them, it is convenient to introduce 
the complex wave $\bar \phi_{\mathbb C}$, such that $\bar \phi = 2\Re (\bar \phi_{\mathbb C})$ 
in \eq{awp}. At late time, using the {\it out} mode basis and \eq{Smat}, one finds 
\be
\bar \phi_{\mathbb C}(x,t\to +\infty) = \bar \phi_{\mathbb C}^+(t,x) + \bar \phi_{\mathbb C}^-(t,x).
\label{phiC}
\ee
The two complex waves are 
\bsub \bea
\bar \phi_{\mathbb C}^+(x,t\to +\infty) &=& \bar N A \int \alpha_\om e^{- \frac{(\om - \bom)^2}{2\sigma_0^2 \bom^2}} \frac{e^{-i (\om t - k_\om x)}}{\sqrt{4\pi \vert  \Om_{\rm out} v_g^{\rm out}\vert }} \frac{d\om}{n_\sigma^{1/2}} , \\
\bar \phi_{\mathbb C}^-(x,t\to +\infty) &=& \bar N A \int \beta_\om e^{- \frac{(\om - \bom)^2}{2\sigma_0^2 \bom^2}} \frac{e^{-i (\om t + k_{-\om} x)}}{\sqrt{4\pi \vert \Om_{\rm out} v_g^{\rm out}\vert }} \frac{d\om}{n_\sigma^{1/2}} .
\eea \esub
where $\Om_{\rm out}= \Omega(k_\om^{\rm out})$ is the co-moving frequency of the outgoing modes, 
and $v_g^{\rm out}$ their group velocity. 
Evaluated through a saddle point method, one obtains 
\bsub \label{2refl}\bea
\bar \phi_{\mathbb C}^+(x, t\to +\infty) &\sim&  A_{\bar \om} \times  \alpha_{\bom}\, \varphi_{\bom}^{\rm out}(x) e^{-i \bom t} \times e^{- \frac{\sigma_0^2 \bom^2}2 (t- x/v_g^{\rm out})^2} ,\\
\bar \phi_{\mathbb C}^-(x, t\to +\infty) &\sim& A_{\bar \om} \times  \beta_{\bom}\, (\varphi_{-\bom}^{\rm out}(x))^* e^{-i \bom t} \times e^{- \frac{\sigma_0^2\bom^2}2 (t- x/ v_g^{\rm out})^2} ,
\eea \esub
where their common amplitude factor $A_{\bar \om}$ is given by 
\be
A_{\bar \om} = A \sqrt{2 \sigma_0 \bom \pi^{1/2}}. 
\label{Abom}
\ee
We now have all the ingredients to consider the limit $\bom \to 0$, keeping $\sigma_0$ constant.
Using \eq{Bogodiv}, and 
\be
\alpha_{\om} \sim |\alpha_{\om}| e^{i (\theta+\theta')} , 
\quad \beta_{\om} \sim |\alpha_{\om}| e^{i (\theta' - \theta)},
\label{Bogophase}
\ee
to characterize the phase of the coefficients,
we see that the two outgoing waves of \eq{2refl} merge with each other and 
give a (real) zero-frequency wave of fixed profile. Indeed, using
\be
\alpha_{\bom} \, \varphi_{\bom}^{\rm out}(x) e^{-i \bom t} + \beta_{\bom} \, (\varphi_{-\bom}^{\rm out}(x))^* e^{-i \bom t} \underset{\bom \to 0}{\sim} 2 |\alpha_{\bom}| e^{i(\delta + \theta')} 
\times e^{-i\bom (t-x/v_g^Z)} \Phi_U(x), 
\label{merge}
\ee 
where
\be
\Phi_U(x) \doteq \Re\left\{e^{i\theta} \phi^{\rm out}_0(x)\right\} . \label{Uprofile}
\ee
Using the limit $\om \to 0 $ of \eq{merge}, we get 
\be
\Phi_U(x) = \bar N \frac{\cos\left(k_Z x + \theta \right)}{\sqrt{4\pi v_\as k_Z v_g^Z}}. \label{Uprofileas}
\ee
We see that $\Phi_U(x)$ is independent of $t$, $\bom$, and $\delta$. 
Taking the real part of \eq{merge}, we see that the outgoing real wave factorizes and takes the simple form 
\be
\bar \phi(x,t\to +\infty) \sim 4 \left|A_{\bom} \alpha_{\bom}\right|\times 
\Phi_U(x) \times \cos(\bom(t-x/v_g^Z) - \delta - \theta') \, 
e^{-\frac{\sigma_0^2 \bom^2}2 (t - x/v_g^Z)^2}.
\label{outU}
\ee
From this expression, we see that $\Phi_U(x)$ gives the profile of the undulation of $\phi$, 
and that the cosines furnishes a long wavelength modulation\footnote{This factor was absent the first version of this work. We are grateful to Iacopo Carusotto to have pointed out its presence. This modulation was observed and discussed in the Sec.~3.1 of~\cite{Mayoral11}. }.
Since this slow modulation becomes a constant when $\bom \to 0$, 
the outcome is a standing wave described by $\Phi_U(x)$, with nodes at fixed locations. 
This is our first important result. 

To discuss this wave in physical terms, we compute the corresponding fluctuation of the free surface.
Using \eq{deltah} and \eq{Uprofileas}, one finds 
\be
\delta \bar h^{\rm out}(x) \sim \delta \bar h^{\rm out} \sin\left(k_Z x + \theta \right) \times 
\cos(\bom(t-x/v_g^Z) - \delta - \theta') \, e^{-\frac{\sigma_0^2 \bom^2}2 (t - x/v_g^Z)^2}, 
\label{outUndul} 
\ee
where the amplitude is
\be
\delta \bar h^{\rm out} = h_\as \bar N \sqrt{\frac{4\Om_{\rm out}}{\pi v_g^Z c_\as^4 }} |A_{\bom} \alpha_{\bom}| .\label{outUndulAmpl}
\ee
To get rid of the dependence on the amplitude $|A|$ and the normalization $\bar N$, we compute the amplification factor. 
Using Eqs.~\eqref{indeltahAmpl} and \eqref{outUndulAmpl}, we find 
\be
\frac{\delta \bar h^{\rm out}}{\delta \bar h^{\rm in}} = \frac{(c_\as - v_\as)}{\bom} \sqrt{\frac{ v_\as k_Z \bar \kappa}{2 \pi c_\as v_g^Z}} .\label{amplfactor}
\ee
Irrespectively of any choice, the ratio diverges as $1/\bom$ for $\bom \to 0$. This is our second important result.

When sending \emph{several} low frequency wave packets, 
each characterized by its own phase $\delta_i$, the 
outgoing waves \eq{outUndul} will all have in common the {\it same} short wavelength 
profile characterized by the $\sin\left(k_Z x + \theta \right) $. Hence, the undulation profile 
is insensitive to initial phases $\delta_i$. Instead, the undulation amplitude does depend on them since it given by a sum containing the $ \cos(\bom(t-x/v_g^Z) - \delta_i - \theta')$. 
As we shall see in the next section, when taking into account 
the low frequency noise that would be present in every experiment,
this implies that in the linear regime
the amplitude of the undulation is, in effect, unpredictable. 
Before studying this important fact, we make three extra comments that lead to specific predictions which could hopefully also 
be tested in future experiments. 

First, the linear treatment 
predicts that both signs of the amplitude are equally possible, 
since the sign is governed by the cosine factor in \eq{outUndul},
or by an oscillating sum if several waves are sent. 
This is not the case when working with the forced KdV equation~\cite{Wu87}. 
In addition, this symmetry will also be lost when including non-linear effects. 
This lost has been recently found in a similar context~\cite{Michel13}.

Second, to be more specific, we suppose that $v_\as \ll c_\as$. 
This means that the outgoing reflected waves are deep water waves. 
It simplifies the expressions and is relevant for many experiments. In this case, using 
\eq{kZ} and \eq{HJ}, we get
\be
k_Z = \frac{c_\as^2}{h_\as v_\as^2} \qquad \textrm{and} \qquad v_g^Z = v_\as/2.
\ee
Hence the net amplification factor of \eq{amplfactor} becomes 
\be
\frac{\delta \bar h^{\rm out}_A}{\delta \bar h^{\rm in}_A} = \frac{c_\as^2}{\bom} \sqrt{\frac{\bar \kappa}{\pi h_\as v_\as}} .
\ee
We see that the amplification grows as $1/\sqrt{v_\as}$ for $v_\as \to 0$. This growth will be regulated by capillary effects 
which have not been taken into account in the present analysis.

Third, the phase $\theta$ which governs the location of the asymptotic nodes in 
\eq{outUndul} is complicated because it accounts for the 
propagation from the blocking region to the asymptotic region. 
On the contrary, the undulation profile has a rather universal behavior near the blocking region, 
where the linearized approximation $v_x-c \sim - \kappa x$ holds, 
that is, for $|\kappa x| \ll D$ (see Fig~\ref{vprofile_fig}). 
In addition we restrict our attention to the region where $k_\om \lesssim 1/d_\Lambda$ (i.e., $|\kappa x| \lesssim 1$), 
meaning that one can approximate $F^2(k)$ in \eq{HJ} by $k^2 - d_\Lambda^2 k^4/3$. 
In this region, using the results of 
Appendix \ref{HRApp} and \eq{deltah}, up to an overall constant factor, we get 
\be
\delta h_U(x) \propto \Ai\left( - x/d_{\rm broad} \right), 
\label{Airy}
\ee
where $\Ai$ is the Airy function~\cite{AbramoSteg}, and
where the effective length is given by 
\be
\label{broad}
d_{\rm broad} = \left(\frac{c_0 d_\Lambda^{\, 2}(0)}{6\kappa}\right)^{1/3}. 
\ee In agreement with the results of~\cite{Finazzi10b,Coutant11,Fleurov12}, 
this {\it broadening} length governs the behavior of the undulation 
near the blocking point. This is our third result. 
Notice that $d_{\rm broad}$ depends on three quantities with fractional powers: 
the dispersive length $d_\Lambda(0)$, the gradient $\kappa = \partial_x (v_x - c)$, and the speed $c$, all evaluated at the blocking point. 

We also note that these results also apply to undulations (of density fluctuation) 
found in Bose condensates where the dispersion is anomalous: $F_{\rm BEC}^2 = k^2 + \xi^2 k^4/4$. 
In that case, $d_\Lambda(0) $ is related to the healing length $\xi$ by $d_\Lambda(0) = \sqrt{3}\xi / 2$, 
and the undulation lives where the flow is supercritical, as was verified 
in the experiment of~\cite{Rolley07}.

\subsection{Inclusion of low frequency noise} 
\label{Stoch_Sec}
In realistic conditions, low frequency modes are excited in a non controllable manner,
for instance by the noise of the pump used to create the flow. 
To describe in simple terms the noise, we assume that the incoming waves are described  
by a Gaussian distribution with a power growing like $k_B T/\omega$ for $\om \to 0$. Hence $T$ can be seen as an 
effective temperature. 
Using the field decomposition of \eq{awp} in terms of {\it in} modes, this means
that the coefficients $a_\om$ and $b_\om$ are now treated as random variables~\cite{VanKampen}, 
with the following statistical moments 
\bsub \bea
\mev{a_\om} = 0 , &&\quad \mev{a_{\om'}^* a_\om} = n_\om^{a} \delta(\om - \om'),\\
\mev{b_\om}= 0 , &&\quad \mev{b_{\om'}^* b_\om} = n_\om^{b} \delta(\om - \om').
\eea \label{thstate} \esub
We also assume that the two variables are independent, i.e. $\mev{b_{\om'}^* a_\om}= 0$. 
Since  positive and negative energy 
modes respectively come from the right ($R$) and left ($L$) 
asymptotic regions, they don't share the same effective temperature. 
Moreover, their co-moving frequency $\Om$ 
is Doppler shifted by the right or left asymptotic values of the flow velocity, 
see~\cite{Macher09b} for a discussion about thermal states in fluid flows.
To take both effects into account, we parameterize the low frequency powers as
\bsub \bea
n_\om^{a} &=& \frac{k_B T_R}{N \om},\\ 
n_\om^{b} &=& \frac{k_B T_L}{N \om}.
\eea \label{abspread} \esub
In this state, the average value of the surface perturbation identically vanishes $\mev{\delta h(t,x)} = 0$. 
Indeed, since initial phases 
are random, when averaging over $\delta$ in \eq{outUndul}, 
the mean value vanishes. 
On the other hand, the spread of $\delta h$  is non trivial. 
By a calculation similar to that of \eq{merge}, and 
using Eqs.~\eqref{deltah}, \eqref{awp}, 
\eqref{thstate} one finds\footnote{For more details, we refer to the Chapter 4 of~\cite{Coutantthesis}, which 
presents results from~\cite{Coutant11,Coutant12}.} 
\be
\mev{ (\delta h(t,x))^2 } = 8 \int_0 (n_\om^{a} + n_\om^{b}) |\alpha_\om|^2 d\om \times \left(\delta h_U(x)\right)^2 . 
\label{divUndul}
\ee
This expression establishes that the relative amplitude 
and the position of the nodes are {\it not} affected by 
the randomness of initial conditions. 
On the other hand, its amplitude is a stochastic variable whose spread is fixed by the above expression.  
Therefore, the linearized treatment predicts that 
there is a high probability of observing a macroscopic undulation,
with equal probability to find either sign. This is our fourth result. 

In addition, when integrating over low frequencies, 
the above integral 
diverges since its integrand behaves as $1/\om^2$. 
To regulate it, we consider a flow which has been formed for a finite amount of time $t$. 
This effectively introduces a low frequency cut-off $\sim 1/t$ in \eq{divUndul}, and gives 
\be
\int_{1/t} (n_\om^{a} + n_\om^{b}) |\alpha_\om|^2 d\om \sim \frac{k_B (T_R+T_L)}{2\pi N} \times \bar \kappa t.
\ee
We see that  the diverging character 
for low frequencies engenders a linear growth in time. 
(This result has been confirmed by numerical simulations in atomic Bose condensates~\cite{Mayoral11}.)
We also see that the low frequency waves coming from the right and the left both contribute,
with their respective powers given in \eq{abspread}. Of course, the growth 
will ultimately saturate due to nonlinearities, dissipation, or an infrared cut-off as in the case of transverse modes~\cite{Coutant12}.

\section{Conclusions}

In this paper we studied the scattering of low frequency waves in supercritical flows when the stationary free surface is flat. 
In the zero-frequency limit, we showed that the scattering possesses very specific 
properties. First, the two reflected waves of opposite energy merge and form a single wave with a fixed spatial profile 
and nodes at specific places, see \eq{merge}. Second, the amplification factor relating the amplitude 
of the incoming and outgoing waves diverges as the inverse of the conserved frequency, see~\eq{amplfactor}. 
Third, this factor also depends on a combination of initial phases. 
When considering several wave packets, the outgoing waves 
interfere, and affect the undulation amplitude but not its spatial profile. 
Fourth, near the blocking point, this profile is given by an Airy function
governed by a composite length scale formed with the dispersive length $d_\Lambda$ 
and the gradient of the flow $\kappa$, 
see \eq{broad}.

These properties tell us that free surfaces (which contain no undulation) are unstable when sending low frequency incoming waves,
and that the large and unpredictable amplitude of the undulation is an expression of this instability.
This is confirmed when taking into account the low frequency noise that would 
inevitably present in any flume.
In agreement with~\cite{Mayoral11,Coutant11,Coutant12}, we found in \eq{divUndul} that the undulation 
spatial profile is not subject to any randomness while its amplitude is a random quantity.
In addition, the spread of this amplitude diverges for very low frequencies. 
This divergence is regulated when considering that the stationary flow only existed for a finite lapse of time. The linear treatment predicts a growth of the squared amplitude 
which is linear in this lapse when the incident noise diverges as $1/\om$, see \eq{abspread}. 
It would be interesting to experimentally test this prediction, as that concerning the profile 
of the undulation given in \eq{Airy}. 

As we saw in \eq{outU}, the first effect due to the non-vanishing character
of the mean frequency of the incident wave is a long wavelength modulation of the undulation. This modulation describes 
the intermediate regime which 
interpolates between the zero frequency limit 
which produces a single stationary wave described $\Phi_U(x)$ of \eq{Uprofileas}, 
and the usual scattering at higher frequencies, 
where two distinct wave packets with different respective weights
propagate away from each other with different group velocities. 
In a future work, we hope to describe in more detail 
this intermediate regime. In addition, the low frequency divergence of \eq{Bogodiv} 
should be regulated when including nonlinear effects. 
In this respect it would be particularly interesting to understand the transition from 
a growing undulation whose amplitude is a random variable to a saturated amplitude which is deterministically fixed by nonlinear equations. This question is relevant for both classical fluids and quantum ones, such as dilute atomic Bose gazes, see~\cite{Michel13} for a first study. 

\acknowledgements{We are grateful to Germain Rousseaux and Florent Michel for 
suggestions and interesting remarks. We also thank Iacopo Carusotto for his remark concerning the long wavelength modulation of \eq{outU}.
A.C. would like to thank Baptiste Darbois-Texier, as well as the LadHyX (Ecole Polytechnique, France) 
for their greeting and interest when this work was presented in october 2012.}

\appendix
\section{Dynamics of 2-dimensional surface waves}
\label{UnruhApp}

In this appendix, we consider the propagation of surface waves in the presence of both a current and an uneven bottom.
Following the recent treatment of~\cite{Unruh12}, 
we first derive a non-linear equation for the surface, when the flow is stationary. 
As often done in two-dimensional problems~\cite{Lamb40}, we use the potential and stream function as a new pair of coordinates. 
This hodograph transformation allows us to map the uneven shape of the water flow into a rectangular strip. 
Then, we obtain the equation for the linear perturbations on a background solution. 
Notice that we shall not solve the non-linear equation for the background. 
Rather we show that, by an appropriate choice of the bottom, 
one can obtain a super-critical background flow with a flat surface, i.e. without undulation. 
As a result, the (linear) perturbations are not forced but freely propagate. 

A significant difficulty comes from the fact that we cannot work in the
standard slowly varying approximation where the wavelength is assumed to be smaller
that the typical length characterizing the background variation. 
We shall thus carefully derive the wave equation without any short wavelength
approximation. To complete this presentation, we briefly compare our equation with 
standard approaches~\cite{Johnson,Mei}. 

As a last step, for the interested readers, we show that in the shallow-water wave limit, surface perturbations propagate as a relativistic field on a curved space-time metric. 
This leads to the notion of acoustic black hole~\cite{Unruh81,Balbinot06}, and to experiments aiming at detecting the analog of the Hawking effect.

\subsection{Setup}

The fluid is assumed to be inviscid, incompressible, irrotational, and in a constant gravitational field $\vect g = g \vect{e_y}$. In this case, 
it is well-known that the Navier-Stokes equation and the continuity equations simplify. 
To proceed we define the velocity potential $\vect v = \vect{\grad} (\Phi)$ which satisfies the Laplace equation
\be
\Delta \Phi = 0. \label{Laplace}
\ee
Moreover, the pressure field is obtained through the Bernouilli equation
\be
\p_t \Phi + \frac12 v^2 + gy + p(x,y,z) = 0. \label{Bernouilli} 
\ee
We now also assume that the flow in 2-dimensional. Hence the above quantities  
only depend on the cartesian coordinate 
$y$ (height) and $x$ (longitudinal direction) but not on the transverse direction $z$. The main advantages is that in 2 dimensions, the Laplace equation possesses interesting mathematical properties related to its conformal invariance. 

To study the dynamics of a free surface, one needs to consider 
the boundary conditions for Eqs. \eqref{Laplace} and \eqref{Bernouilli}. 
In our case, we have a one-dimensional water tank, with a given (static) 
profile on the bottom at $y = \zeta_B(x)$. 
At $\zeta_B$, the velocity component orthogonal to the bottom vanishes, i.e. 
\be
v_{y | y=\zeta_B} - v_{x|y=\zeta_B} \p_x \zeta_B = 0,
\ee
where $v_y$ and $v_x$ are the vertical and longitudinal components of $\vec v$, here evaluated
along the bottom surface. At the free surface $\zeta_S(x,t)$, we also have the  
condition that no flux goes across it 
\be
v_{y | y=\zeta_S} - \p_t \zeta_S(x,t) - v_{x|y=\zeta_S} \p_x \zeta_S = 0, \label{BCS}
\ee
where the second term accounts for the time dependence of the free surface. 
The other difference is that here $\zeta_S$ is an unknown function. Therefore, we  
need an extra boundary condition which in our case states that the surface is unconstrained. 
When capillary effects are neglected, it is given by the fact that the pressure is constant on the surface, equal to the atmospheric one $p_0$. Hence, $\zeta_S$ also obeys
\be
\p_t \Phi_{y=\zeta_S} + \frac12 v^2_{y=\zeta_S} + g\zeta_S = 0, \label{BCfree}
\ee
where $p_0$ has been absorbed in the definition of $\Phi$.
These boundary conditions are quite complicated 
because they depend on the value of the unknown potential function evaluated at the unknown position of the free surface, i.e., the dynamical quantities act both as function and arguments of functions. However, in 2 dimensions, this can be circumvented by using an appropriate  
set of coordinate, and by treating the cartesian coordinates $x,y$ as functions.

\subsection{Appropriate coordinates}

We define the stream function $\Psi$ by the relation 
\be
\vect{\nabla} \Psi = \vect{e_z} \wedge \vect v.
\ee
An alternative way is to build $\Psi$ such that $\Phi - i \Psi$ is an holomorphic function of $x+iy$. The 2 potential functions $\Phi(x,y)$ and $\Psi(x,y)$  
satisfy 
\bsub \bea
\vect{\nabla} \Phi . \vect{\nabla} \Psi &= &0, \label{phipsirel1} \\
\vect{\nabla} \Phi . \vect{\nabla} \Phi &=& v^2 , \label{phipsirel2} \\
\vect{\nabla} \Psi . \vect{\nabla} \Psi &=& v^2 ,
\eea \label{confphipsi} \esub
and are both harmonic functions. 
The idea is to use $\Phi$ and $\Psi$ as new coordinates, and $x$ and $y$ as  
unknown functions. To this end, we assume that the velocity flow is nowhere vanishing. 
In that case, the Eqs.~\eqref{confphipsi} guarantee that the mapping $(x,y) \mapsto (\Phi,\Psi)$ is 
a diffeomorphism, and defines the cartesian {\it functions} $\hat x$ and $\hat y$ by the relation 
\bsub \bea
\hat x\left(\Phi(x,y), \Psi(x,y) \right) &=& x, \label{hatx}\\
\hat y\left(\Phi(x,y), \Psi(x,y) \right) &=& y. \label{haty}
\eea \esub
This means that $(\hat x,\hat y)$ is the reciprocal function of $(\Phi,\Psi)$. To distinguish functions from variables, we note $(\phi,\psi)$ (instead of $(\Phi,\Psi)$) the new coordinate set. 
Moreover, their partial derivatives are related by 
\be
\bmat \p_\phi \hat x & \p_\psi \hat x \\ \p_\phi \hat y & \p_\psi \hat y \emat = \frac1{v^2} \bmat \p_y \psi & - \p_y \phi \\ -\p_x \psi & \p_x \phi \emat . \label{coordchange}
\ee
Using Eqs.~\eqref{confphipsi}, a straightforward computation shows that  
$\hat x, \hat y$ are harmonic, i.e.
\be
\p_\phi^2 \hat x + \p_\psi^2 \hat x = \Delta_{(\phi,\psi)} \hat x = 0,
\ee
and similarly for $\hat y$. The main interest of this new coordinate set is that surfaces of constant $\Psi$ are streamline. Therefore, for {\it stationary} flows, 
the bottom and the surface are both located at constant $\Psi$. By convention, we set the bottom at $\Psi=0$, hence the boundary condition reads 
\be
\hat y(\phi,0) = \zeta_B(\hat x(\phi,0)).
\ee
Similarly, at the surface, at $\psi = \psi_S$ we have \eq{BCfree} in the new coordinate set\footnote{Notice that since we assume a stationary flow, the constant term of the Bernoulli equation cannot be absorbed in the time derivative of the potential. Here we determined it using the knowledge of the asymptotic value of the velocity and water height.}\textsuperscript{,}\footnote{Using \eq{coordchange}, one can alternatively write this equation with the unknown function $\hat x$ only.} 
\be
\frac1{2 \left((\p_\phi \hat y)^2 + (\p_\psi \hat y)^2 \right)} + g \hat y = \textrm{const} = \frac12 v_\as^2 + gh_\as. \label{backgroundequ}
\ee
This equation is still a complicated nonlinear differential equation, and its resolution might be rather involved. However, this equation is quite convenient to solve the ``inverse problem'', namely, if one chooses the shape of the free surface $\hat y(\phi,\psi_S)$, one can easily determine the profile of the bottom from the latter equations, as done in~\cite{Unruh12}. 
In addition, this description will turn out to be very efficient to derive the wave equation for linear perturbations.

\subsection{Linear perturbations}

We now study linear perturbations on top of a stationary solution $\Phi_0(x,y)$ of the preceding set of equations.
In other words, we study the free (un-forced) solutions of the form $\Phi(x,y;t) = \Phi_0(x,y) + \delta \Phi(x,y;t)$ to first order in $\delta \Phi$.  
The corresponding perturbation of the velocity flow is $\delta \vect v = \vect \nabla_{(x,y)} \delta \Phi$. 
(To enlighten the notations, we shall refer to the background flow as $\vect v = v_x \vect{e_x} + v_y \vect{e_y}$, without $0$-index.) 
Because the location of the free surface also changes, we must perform two linear expansions, one for the functions, and the other for the argument. 
Hence, we shall keep explicitly the zeroth order quantities. 
As a first step, we write Bernouilli equation \eqref{Bernouilli} at first order 
\be
\frac12 v^2(x,y) + gy + \p_t \delta \Phi(x,y;t) + \vect{v} . \vect \nabla \delta \Phi(x,y;t) + p(x,y;t) = \frac12 v_\as^2 + gh_\as + p_0. 
\ee 
The pressure term is given by $p = p_0 + \delta p$, where $\delta p$ is small and vanishes at the free surface. 
As above, the location of the free surface is best expressed in terms of the abstract set of coordinates $(\phi, \psi)$. 
We must be cautious here: we shall use the background potentials as coordinates, not the exact ones. 
This means that we change functions of cartesian coordinates as 
\be
f(x,y;t) \to f\left(\hat x_0(\phi,\psi), \hat y_0(\phi,\psi); t \right).
\ee
In these coordinates, we have $\vect{v}. \vect \nabla = v^2 \p_\phi$. Since the exact flow is not stationary, the free surface is no longer characterized by a constant value of $\psi$. 
To first order it is described by 
\be
\psi = \psi_S + \delta \Psi_S(\phi,t). \label{perturbsurf}
\ee
Hence the vanishing pressure change $\delta p$ at the free surface gives  
\be
\left[ \frac12 v^2(x,y) + gy + \p_t \delta \Phi(x,y;t) + \vect v . \vect \nabla \delta \Phi(x,y;t) \right]_{\psi = \psi_S + \delta \Psi_S} = \frac12 v_\as^2 + gh_\as. 
\ee 
Since the background flow satisfy the zeroth order equation, the remaining terms give 
\be
\Big[ (\p_t + v^2 \p_\phi) \delta \Phi + \underbrace{\p_t \p_\psi \phi \delta \Psi_S}_{= 0} + G \delta \Psi_S \Big]_{\psi = \psi_S} = 0, \label{pertubBernS}
\ee 
where $G = \p_\psi (g\hat y_0 + v^2/2)$. A few calculations using Eqs.~\eqref{BCS} and \eqref{BCfree} give
\be
v_x G = g +\frac{v_x^2}{v^2} \p_y(v^2/2) . \label{vxG}
\ee
To determine $\delta \Psi_S$, we know that \eq{perturbsurf} holds when taking the Lagrangian time derivative $D_t = \p_t + \vect v. \vect \nabla + \delta \vect v . \vect \nabla$. This gives the equation  
\be
(\p_t + v^2 \p_\phi) \delta \Psi_S = v^2 \p_\psi \delta \Phi.
\ee
Therefore, applying $(\p_t + v^2 \p_\phi)$ to \eq{pertubBernS}, we obtain 
\be
(\p_t + v^2 \p_\phi)\frac1G(\p_t + v^2 \p_\phi) \delta \Phi +  v^2 \p_\psi \delta \Phi = 0.
\ee 
The last step is to relate $\p_\psi \delta \Phi$ to $\p_\phi \delta \Phi$. To this aim, we shall use the standard method~\cite{Johnson}, i.e., integrate the harmonic equation in the volume of the fluid, and use the bottom boundary condition. The bottom is still characterized by $\Psi = 0$ and thus $\delta \Psi(\phi, 0) = 0$. Moreover, using \eq{phipsirel1} at first order, we have 
\be
v^2 \p_\phi \delta \Psi = - v^2 \p_\psi \delta \Phi.
\ee
In particular, in the bottom, $\p_\psi \delta \Phi = 0$. Therefore, in Fourier transform, we solve the harmonic equation for $\delta \Phi$
\be
\delta \Phi(\phi,\psi;t) = \int A(k,t) e^{i k \phi} \cosh(k\psi) dk.
\ee
From this, we deduce at the surface
\be
\p_\psi \delta \Phi = -i\p_\phi \tanh(-i\psi_S \p_\phi ) \delta \Phi.
\ee
This gives us the wave equation for surface waves over arbitrary bottoms 
\be
(\p_t + v^2 \p_\phi)\frac1{G}(\p_t + v^2 \p_\phi) \delta \Phi -i v^2 \p_\phi \tanh(-i\psi_S \p_\phi ) \delta \Phi = 0. 
\ee 
To obtain the equation of the body of the paper, we shall use the $x$ coordinate instead of $\phi$. This does not mean that we go back to cartesian coordinate set $(x,y)$. Rather it means that we use a mixed set $(x,\psi)$, so that first, the longitudinal coordinate has its usual physical interpretation, and second  
the (background) free surface is still simply characterized by $\psi = \psi_S$. 
Hence, along the background free surface, using $v^2\p_\phi = v_x \p_x$, we get 
\be
(\p_t + \p_x v_x)\frac1{v_x G}(\p_t + v_x \p_x) \delta \Phi -i\p_x \tanh \left(-i\frac{\psi_S v_x}{v^2} \p_x \right) \delta \Phi = 0, \label{UnruhwavequApp} 
\ee 
where we divided by $v_x$ in order to change the ordering of $\p_x$ and $v_x$ in the first parenthesis, 
and to obtain a self-adjoint wave operator. 

We here note that the value of $\psi_S$ can be related to the (cartesian) water depth $h_B \equiv \zeta_S - \zeta_B$ at fixed $x$. Indeed, from \eq{coordchange}, we have 
\be
h_B(x) = \int_0^{\psi_S} \frac{v_x}{v^2} 
d\Psi.
\ee
Asymptotically, this means
\be
h_\as = \frac{\psi_S}{v_\as}. \label{psiS}
\ee
In addition, to lowest order in the (vertical) gradient $\partial_\Psi$, the depth $h_B(x)$ reduces to 
\be
h_B(x) \simeq h_\as \frac{v_x v_\as}{v^2} \equiv d_\Lambda(x). \label{WKBdepth}
\ee
To conclude, we relate $\delta \Phi$ to the vertical 
variation of the free surface with respect to the background one $\delta h(x,t) = \zeta_S(x,t) - \zeta_{S0}(x)$. By definition of the free surface in \eq{perturbsurf} (remember that the $\psi$ there is the background stream function) we have
\be
\Psi(x, \, \zeta_{S0}(x) + \delta h(x,t) ) = \psi_S + \delta \Psi_S \Big(\Phi_0 \big(x,\zeta_{S0}(x)+ \delta h(x,t) \big) ;t\Big). 
\ee
At first order in $\delta h$, this gives
\be
\underbrace{\p_y\Psi_0}_{= v_x} \delta h = \delta \Psi_S. \label{Appdeltah}
\ee
From \eq{pertubBernS} in the mixed coordinate set $(x,\psi)$, we derive the relation $\delta \Psi_S = (1/G) (\p_t + v_x \p_x)\delta \Phi$. 
Combining it with \eq{Appdeltah} and the expression for $v_x G$ of \eq{vxG}, we get \eq{deltah}.
In the body of the paper, we enlighten the notations by writing $\phi$ instead of $\delta \Phi$ 
and use the dispersive scale $d_\Lambda(x) \equiv \psi_S v_x/v^2$. Using \eq{psiS}, $d_\Lambda$ reduces to the expression used in \eq{dLambda}.

\subsection{Validity conditions}
\label{valid_App}

We now briefly discuss the validity of the key equations, i.e. the background equation 
\eqref{backgroundequ} and the linear wave equation \eqref{UnruhwavequApp}. 

Our wave equation \eqref{UnruhwavequApp} describes the propagation of linearized waves on 
the top of a inhomogeneous background flow which is due to a current above an uneven bottom. 
In standard treatments, inhomogeneities are assumed to be ``slowly varying''~\cite{Johnson,Mei,Dingemans}, 
or the current is neglected in order to consider appreciable variations
of the water height~\cite{Mei}.\footnote{For the interested reader, we here explain with more details why the standard treatments are inadequate to compute the scattering coefficients in the flow we considered. We use the treaty of C.C. Mei~\cite{Mei} to explain the situation. In chapter 3, currents and varying bottoms are both considered (see in particular Sec. 3.6), but they are assumed to be ``slowly varying'', and therefore the physical predictions are limited to the regime of ``ray approximation''. As we explained, 
when using this (WKB) approximation, the "over-reflection", i.e. the mixing amongst modes of opposite norms is automatically neglected. Hence this treatment cannot be used to derive the key equation \eqref{HRrat}. In chapter 4, the author considers bottoms with appreciable variations, and hence the scattering of long wavelengths can be, and in fact is, studied in details. However, in this treatment, the crucial roles played by the current (and especially when it becomes supercritical) are simply not considered. In particular, the role of ``negative energy waves'' which are necessary to obtain an "over-reflection", is simply not discussed. When consulting other treaties, such as~\cite{Johnson,Dingemans}, we meet the same situation: there is no treatment allowing the description of the scattering of long wavelength modes in inhomogeneous super-critical flows with non-negligible spatial gradients.} 
Importantly, \eq{UnruhwavequApp} is not based of any kind of WKB approximations, 
which means that it applies to modes with arbitrary long wavelength. 
This is essential for the present paper since the peculiar aspects of the scattering we studied are (only) found in the zero frequency limit. 
This is to be contrasted to the standard description of wave blocking involving an Airy function. 
As shown in~\cite{Coutant14}, the later becomes valid precisely when the 
mode amplification we studied disappears.

In addition, the non-linear equation for the background is not restricted to a ``weakly non-linear'' regime. 
Rather it applies only to {\it stationary} backgrounds. 
This is in contrast with descriptions using Boussinesq or Korteweg-de Vries type of equations~\cite{Johnson,Wu87}. Lastly, we did not assume that the flow is ``near critical'' ($F_n \sim 1$). 
The maximum value of the Froude number is only restricted by the requirement that the flow stays non-turbulent. According to e.g.~\cite{Johnson}, this is the case if $F \lesssim 1.2$. The description of the undulation we obtained is thus valid for flows with $F$ from $1 < F \lesssim 1.2$.

\subsection{Link with Relativity}
\label{Relat_App}

A remarkable fact of \eq{UnruhwavequApp}, 
which is the root of the notion of ``acoustic black hole''~\cite{Unruh81}, is its close relationship with the propagation of a  relativistic field in a curved space-time. Explicitly, in the hydrodynamical regime (when $\tanh(k) \sim k$) \eq{UnruhwavequApp} reads 
\be
[(\p_t + \p_x v_x)\frac{\rho}{c^2}(\p_t + v_x \p_x) -\p_x \rho \p_x] \delta \Phi = 0, \label{gwfullEq}
\ee
where we used the functions defined in \eq{gwexpressions}. Written under this form, this equation is identical to that of sound waves in a moving fluid~\cite{Balbinot06,Barcelo05}. 
More remarkably, it also {\it coincides} with the d'Alembert equation of a scalar field in a space-time described by the metric 
\be
ds^2 = \frac{\rho(x)}{c(x)} \left[ c^2(x) dt^2 - (dx - v_x(x) dt)^2 - dy^2 - dz^2\right], \label{ds2}
\ee
when assuming that the field does not depend on $y$ and $z$. 

The relationship  between flows that becomes supercritical and black hole geometries
is then straightforward: In a stationary flow, whenever $c^2$ crosses $v_x^2$, the associated 
metric possesses a black hole (or white hole) horizon.
For more explanations, we refer to~\cite{Jacobson12,Balbinot06}.

\section{Inner scalar product and sign of energy}
\label{energyApp}

The conserved product canonically associated with \eq{Sact}, and \eq{twaveeq}, plays many roles. 
For instance, it governs the anomalous sign in \eq{U11rel}, 
and the notions of completeness and orthogonality of the stationary modes
used to build the wave packets in \eq{awp}. 
For any pair $\phi_1,\phi_2$ of complex solutions of \eq{twaveeq}, it is given by 
\be
(\phi_1\vert \phi_2) = i \int  (\phi_1^{*} \pi_2^{\phantom *} - \pi_1^{*} \phi_2^{\phantom *}) dx,
\label{KGn}
\ee
where we used \eq{H1}
to define the momenta $\pi_1$ and $\pi_2$ associated with $\phi_1$ and $\phi_2$. 
Several important properties should be mentioned. 

First, it is constant, in virtue of Hamilton's equations. 
Second, the norm $(\phi_1\vert \phi_1)$ of any complex solution $\phi_1(t,x)$ 
is the opposite of the norm of its complex conjugated $\phi_1(t,x)^*$. 
In the mathematical literature, this is called a {\it Krein} scalar product~\cite{Bognar}. 
It implies that the norm of the real solutions $\bar \phi(t,x)$ of \eq{awp} always vanishes.
At first sight, this seems to imply that it would play no role in hydrodynamics. As we shall see, this is not the case.

Third, when considering two stationary (complex) modes,
\eq{KGn} vanishes when $\om_1 \neq \om_2$. This guarantees that the 
$\om$-sectors do not mix with each other, and can thus be studied separately. 
To form a complete basis, it is appropriate to separate the modes 
of positive norm from those of negative norm. 
The latter are then given by the complex conjugated of the former. 
Irrespectively of the sign of $\om$, 
$e^{-i \om t} \phi_\om^{\rm in }$ has a positive norm, whereas
$(e^{i \om t}\phi_{-\om}^{\rm in })^*$ has a negative norm.  
The mode basis can thus be taken orthonormal, with all {\it positive norm modes} obeying
\bea
( (e^{-i \om t} \phi_{\om \phantom{'}}^{\rm in }) | (e^{-i \om' t}\phi_{\om'}^{\rm in })) = N \, \delta(\om - \om') ,\label{orthorel} 
\eea
where $\delta(\om - \om')$ is the Dirac distribution 
(because the domain of $x$ is the entire real axis), 
and where $N$ is an arbitrary (real and positive) 
constant which has the dimension of an action. A possible choice for gravity waves which depends on the flow properties is 
\be
N_{\rm gw} = \rho_0^{3D} L_\perp \times h_\as^3 c_\as = \frac{\rho_0^{3D} L_\perp}{g \bar N^2}. 
\label{N_def}
\ee
When using \eq{KGn} and \eq{H1}, one verifies that 
$\bar N = h_\as c_\as^{3/2}$ is the `net' amplitude 
of the modes, which guarantees that they are normalized as in \eq{orthorel}. 
We can then normalize the asymptotic plane waves $\varphi_\om^j$ and $(\varphi_{-\om}^j)^{*}$ associated with the roots discussed in Sec.\ref{NEW_Sec}. 
They are given by
\bsub \bea
\varphi^j_\om &=& \bar N  \frac{e^{i k^j_\om x}}{\sqrt{4\pi \vert \Om(k_\om^j) v^j_g \vert }}, \\
(\varphi^j_{-\om})^* &=& \bar N \frac{e^{-i k^j_{-\om} x}}{\sqrt{4\pi \vert \Om(k_\om^j) v^j_g \vert }}. 
\eea \label{plw} 
\esub
where the superscript $j$ stands for {\it in} or {\it out}. 
Notice also that we use the symbol $\varphi$ to designate the asymptotic plane waves,  
whereas $\phi$ designates the corresponding globally defined solution.~\footnote{
On the right of the blocking point, the four roots have been already discussed. 
On the left, since the flow is supercritical, only two real roots exist. (In total, there are thus six real roots. They are associated with the three $in$ and three $out$ modes~\cite{Macher09}.)
On the usual branch with $\Om > 0$, one finds $k_\om^{\rm co,\, in}$, the incoming co-propagating mode, 
and on the negative $\Omega$ one, there is $-k_{-\om}^{\rm in}$, which describes the incoming mode with negative energy. The two other roots are complex and conjugated to each other. They describe a growing and a decaying mode. 
Since physical modes must be asymptotically bounded~\cite{Greiner,Fulling,Primer}, 
the contribution of the growing mode must vanish.}

Fourth, it is instructive to relate the above mathematical properties to 
the sign of the energy of the wave which has a clear physical meaning.  
For $\om > 0$, one finds that the sign of the norm agrees with that of the energy. 
Hence one can trade one for the other. 
This can be verified by expressing the energy transported by a wave $\bar \phi$ in two different ways:
\be
\bar E = H\left[\bar \phi \right] = \frac12 (\bar \phi | i\p_t \bar \phi).
\ee
In the first equality we used $H$ of \eq{hamilt}, 
and in the second $( \,.\,  \vert\,.\,  )$ is the scalar product of \eq{KGn}. 
Using \eq{awp}, we get 
\be
\bar E = N \int_0^{\om_{\rm max}} \om \left[ {\vert a_\om \vert}^2 - {\vert b_\om \vert}^2 \right] d\om .
\label{Ewp}
\ee
The origin of the minus sign in
\eq{Ewp} can  be viewed as coming from either the negative frequency of the positive norm solution $e^{i \om t}\phi_{-\om}$, or alternatively from negative norm of the positive frequency solution 
$e^{-i \om t}(\phi_{-\om})^*$.
In any case, the real wave ${\rm Re} ( e^{- i \om t} \phi_\om)$ carries a positive energy, 
whereas ${\rm Re}(e^{i \om t}\phi_{-\om})$ carries a negative one. 
Unlike the sign of the frequency and that of the norm which are conventional, 
the sign of the energy is physically unambiguous. 

As a last comment, we wish to point out that 
the scalar product must be used, and has been used in~\cite{Weinfurtner10}, to test, 
from experimental data concerning $\delta h(t,x)$, the validity of the 
Hawking's prediction of \eq{HRbeta}. To this end, one should send a series of monochromatic waves, 
or of broad wave packets as those of Sec.\ref{wp_Sec} described by the real 
profiles $\delta\bar h^{\rm in}(t,x)$. The next step consists in extracting from observational data (by making use of a double Fourier transform in $\om, k$ space) 
the {\it complex} functions $\delta h_{\mathbb C}^{\pm}(t,x)$ 
(i.e. the equivalent of $\phi_{\mathbb C}^\pm$ of \eq{phiC}) describing the positive and the negative energy outgoing waves.
Their norm can then be computed by evaluating \eq{KGn} sufficiently far away from the blocking point where 
the mode mixing has taken place, i.e., so that the WKB approximation applies, see \eq{expli}.
Using \eq{H2} to relate $\delta h_{\mathbb C}^{\pm}$ to $\pi_{\mathbb C}^{\pm}$, and the WKB relation $
\delta h_{\mathbb C}^{\pm}= (i\Omega_{\pm}^{\rm out}/g) \phi_{\mathbb C}^\pm$, 
where $\Omega_{\pm}^{\rm out}(x) = \om \mp v k_{\pm \om}^{\rm out}$,
\eq{KGn} is given by 
\be
(\delta h_{\mathbb C}^{\pm}\vert \delta h_{\mathbb C}^{\pm} ) = 2 g \rho^{3D}_0 L_\perp \int  {|\delta h_{\mathbb C}^{\pm}(x)|^2 \over \Omega_{\pm}^{\rm out}(x)} dx.
\label{KGn2}
\ee
The ratio $(\delta h_{\mathbb C}^-\vert \delta h_{\mathbb C}^{-}) /
(\delta h_{\mathbb C}^+\vert \delta h_{\mathbb C}^{+} )$
must then be compared with the theoretical prediction
of App.\ref{HRApp}, see \eq{HRrat}.

\section{Calculation of the $S$-matrix}
\label{HRApp}

We summarize the essential steps  leading to \eq{HRbeta} in a dispersive medium. 
We follow~\cite{Coutant11} where the interested reader will find a detailed treatment. 
The basic idea is to solve the mode equation \eqref{omwaveeq} at leading order in the quantity $\kappa d_\Lambda /c\ll 1$. This 
means that the flows have low gradients in the units of the dispersive 
length $d_\Lambda$. This condition which should not be confused with the 
standard short wave length approximation which is $\kappa /c \ll k_\om$, 
and which implies $\kappa \ll \omega$.~\footnote{After Unruh's proposal to mimic black hole physics in fluid flows~\cite{Unruh81}, it was emphasized that in such systems, dispersive effects must  
be taken into account~\cite{Jacobson91}. Subsequently, a large amount of (analytical and numerical) work was done to identify under which conditions \eq{HRbeta}
would apply to spectra in dispersive media~\cite{Unruh95,Jacobson93,Brout95,Corley96,Corley97,Unruh04,Balbinot06,Macher09,Finazzi10b,Finazzi12,ScottThesis}. 
It is now clear that the crucial inequality which guarantees small deviations is $\kappa d_\Lambda /c_0 \ll 1$.}

We first simplify \eq{omwaveeq} by replacing 
the centrifugal acceleration, $g_{\rm eff}(x)$, by the constant $g$. 
A first order expansion of \eq{gwexpressions} in $\kappa d_\Lambda /c\ll 1$ shows that it is a legitimate approximation.  
In this case $\rho/c^2$ is a constant, and \eq{omwaveeq} becomes 
\be 
\left[(\om + i\p_x v_x(x) )(\om + i v_x(x) \p_x) - c^2(x) \hat F^2(d_\Lambda(x)) \right] \phi_\om(x) = 0 .
\label{me}
\ee
As we shall see, the non-trivial properties of the scattering  
originate from the region surrounding $x = 0$ where  
the flow becomes super-critical. 
In this region, two additional approximations can be  
implemented. First, the background flow quantities can be expanded to first order in $x$, i.e.,  $v_x \sim c_0 - \kappa_v x$, 
$c \sim c_0 + \kappa_c x$, and thus $d_\Lambda \sim d_0 (1 + 2 \kappa_c x / c_0)$.  
Second, for low frequencies $\omega$, the typical wave numbers are much 
smaller than $1/d_\Lambda$, which means that 
dispersive effects can be described to first order in $d^2_0 k^2$. Using \eq{gwexpressions}, one has 
\be
c^2(x) \hat F^2(d_\Lambda(x)) \sim c_0^2 \, F_4^2 =  -  \partial_x \, c^2(x)  \partial_x - \frac{c_0^2 d_0^2}{3} \partial_x^4 \, . 
\label{Fs}
\ee
These approximations are supported by the fact that the location of the 
blocking point, and $k_{\rm b.p.}(\om)$, the wave number at that point, 
obey $(\kappa/c_0) x_{\rm b.p.}(\om) \ll 1$ and $d_0 k_{\rm b.p.}(\om) \ll 1$~\footnote{When $\om$ is low enough, their precise expressions are $x_{\rm b.p.}(\om) = 1/(2\kappa) (3^{5/6}d_0 \om)^{3/2}$, and $k_{\rm b.p.}(\om) = -(3 \om /d_0^2)^{1/3}$, where $\kappa = \kappa_c + \kappa_v$, see Sec.I.D.1 of~\cite{Coutant11}.}.

Under these assumptions, it is appropriate to solve \eq{me} in Fourier space. Indeed, the Fourier transform $\tilde \phi_\om(k)$ 
obeys a second order equation in $\hat x = i \partial_k $. 
(This is similar to the Airy equation, which is a first order equation in $\hat x = i \partial_k $, from which one immediately obtains
the Fourier transform of the Airy functions.). 
In the present case, the WKB solution has the form~\cite{Coutant11}
\be
\tilde \phi_\om(k) \sim \bar N \sqrt{\frac{\p_\om X_\om(k)}{4\pi c_0 F_4(k,X_\om(k))}} e^{-i \int^k X_\om(k') dk'}, \label{pWKB}
\ee
where $X_\om(k)$ is a $k$-dependent solution of the Hamilton-Jacobi equation at fixed $\om$ 
associated with \eq{me}. Using \eq{Fs}, one gets 
\bsub \bea
(\om - v(X_\om) k)^2 &=& c^2_0 F^2_4(k,X_\om) \\
&=& (c(X_\om) k)^2\, -  \frac{c_0^2 d_0^2}{3} k^4 . \label{pHJ}
\eea \esub
The important fact is that in momentum space, there is no turning point 
(i.e. $\p_\om X = (dk/dt)^{-1}$ stays finite). As a result, the validity of 
\eq{pWKB} is rather easy to handle. In fact, one first verifies that for large momenta 
\eq{pWKB} becomes exact. Hence, the corrections only arise from low momenta. Secondly, for these
momenta, dispersion effects can be neglected (i.e. one can send $d_0 k \to 0$).  
Therefore, the corrections can be evaluated in the hydrodynamical regime, by working with a second order equation in $x$-space. 
These corrections describe the mode mixing between the counter-propagating and co-propagating 
hydrodynamical roots of Fig.~\ref{Disprel_fig}. The evaluation of this mixing goes beyond the scope of the present paper, and is not included in what follows.~\footnote{
It is worth noticing that the linearized Korteweg-de Vries (LKdV) equation also gives the Hamilton-Jacobi equation of \eq{upHJ}. 
Therefore, the same analysis could also be performed starting with the LKdV equation generalized to inhomogeneous background flows. 
However, we decided to work with \eq{twaveeq}
for two main reasons: 
1. to be able to use the conserved scalar product of \eq{KGn} which is canonically associated to \eq{twaveeq}; 
2. to make contact with the relativistic equation \eqref{gwfullEq} used by Hawking. }

When working with \eq{pWKB}, we limit ourselves to the counter-propagating sector of \eq{pHJ}, i.e. $\om - v k = - c_0 F_4$. 
To first order in $d_0^2k^2$, the dispersion relation reads
\be
\om = -\left(c(X_\om) - v(X_\om)\right) k +  \frac{c_0 d_0^2 k^3}{6}. \label{upHJ}
\ee
Near the blocking point, to first order in $X_\om$, the unique solution is  
\be
X_\om(k) = - \frac{\om}{\kappa k} + \frac{c_0 d_0^2 k^2}{6\kappa}. 
\ee
Having computed $X_\om(k)$, we apparently know the WKB mode of \eq{pWKB}. 
However, to fully specify it, one still needs to choose a branch cut for the logarithm appearing in $\int X_\om(k) dk = -\frac{\om}{\kappa} \ln k + \frac{c_0 d_0^2 k^3}{18\kappa}$. 
Interestingly, the two inequivalent choices of the branch cut deliver the two {\it in} modes involved in the $S$-matrix of \eq{Smat}. 

\begin{figure}[!ht]
\begin{center}
\includegraphics[width=0.49\columnwidth]{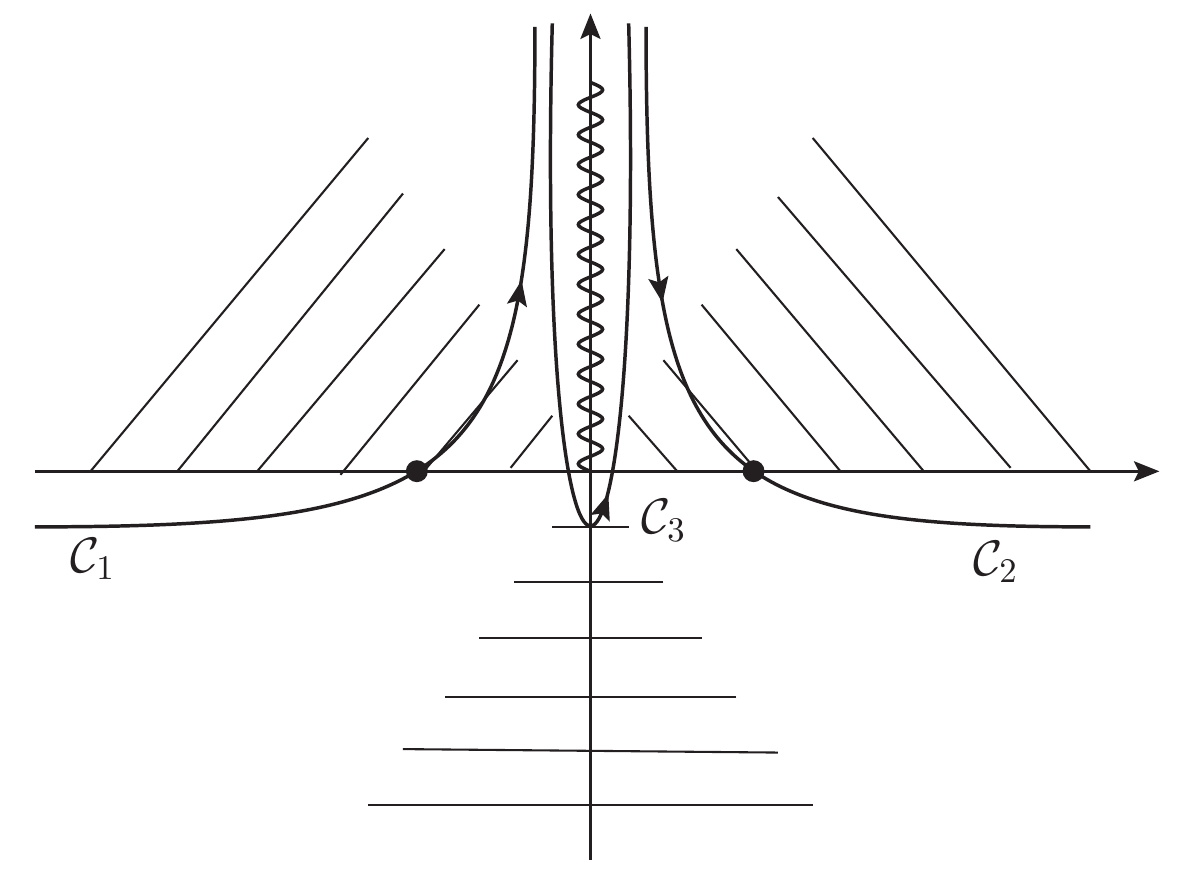} 
\includegraphics[width=0.49\columnwidth]{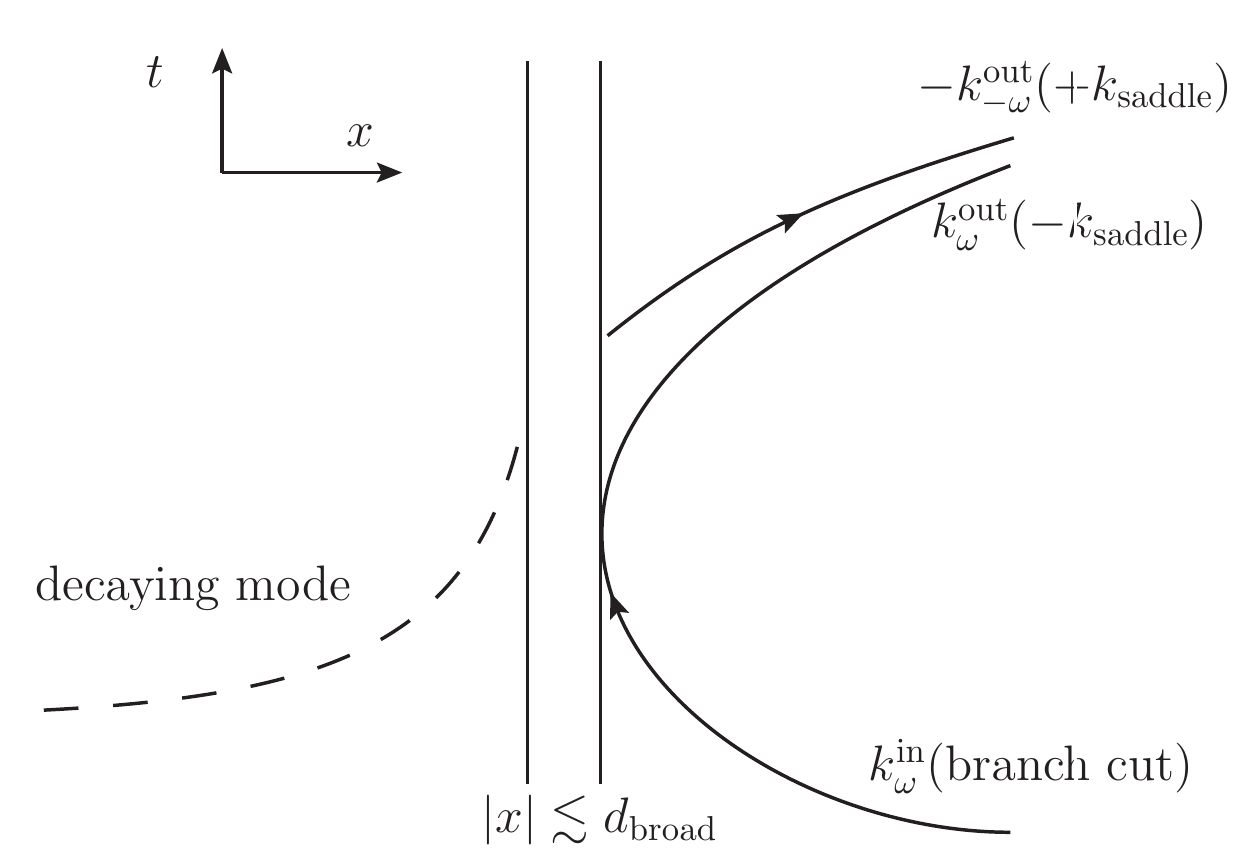} 
\end{center}
\caption{On the left: Contour of integration of \eq{inverseFT} in the complex $k$-plane. The contributions of $\mathcal C_1$ and $\mathcal C_2$ can be evaluated by a saddle point approximation (the locations of the saddles $\pm k_{\rm saddle}$ are indicated by black dots). 
They give the two outgoing modes with high wave number. Instead the contribution of $\mathcal C_3$ is obtained by setting $d_0 \to 0$, and gives the low wavenumber incoming branch. 
The hatched zones represent the directions in the complex plane 
where the integrand of \eq{inverseFT} diverges for $|k|\to \infty$. The contours giving rise to physical
modes will asymptote outside these zones. \\
On the right: the low frequency characteristics of the incoming mode $\phi_\om^{\rm in}$ of \eq{resultB}. They approximatively describe the trajectories followed by the wave packets of Sec.~\ref{wp_Sec}. 
For $|x| \gg d_{\rm broad}$, the mode is well approximated by a superposition of the 3 WKB modes of \eq{expli}, 
the low wave number incoming one, and the two outgoing ones.  
For each mode, we give its root and the origin of its contribution to \eq{inverseFT}. 
See also~\cite{Weinfurtner10} (Fig.~4), where these wave properties have been clearly observed. 
}
\label{Contour_fig}
\end{figure}

Indeed, to obtain 
the positive energy (positive norm) incoming mode $\phi_\om^{\rm in}$, 
the cut should chosen to be such that \eq{pWKB} is analytic in the {\it lower} half-plane, 
because this guarantees that the mode is completely reflected in $x$-space, and vanishes for $x \to - \infty$, see Fig.~\ref{Contour_fig} right panel. 
Had one chosen the mode which is analytic in the upper half-plane, one would have described 
the negative energy incoming mode $(\phi_{-\om}^{\rm in})^*$. 
(Notice that this situation is the white hole (time reversed) version of the 
standard discussion of~\cite{Brout95,Unruh04,Coutant11} which applies to black hole flows.) 
In $k$-space, the incoming mode $\phi_\om^{\rm in}$ is thus described by 
\be
\tilde \phi_\om^{\rm in}(k) =  \bar N \frac{\exp\left[{i \left(\frac\om\kappa \ln (k- i \epsilon) - \frac{c_0 d_0^2 k^3}{18 \kappa} \right)} \right]}{k \sqrt{4\pi \kappa c_0 (e^{\frac{2\pi \om}{\kappa}} - 1)}} 
,  \label{inpmode}
\ee
where the $i \epsilon$ prescription ensures analyticity in the lower half-plane.
We normalized the mode using the scalar product of \eq{KGn} re-expressed in Fourier space, see the Appendix of \cite{Jacobson07b}. 
Notice also that the prefactor of \eq{pWKB} has been here evaluated under the approximation $cF \sim c_0 k$, something valid in the weakly dispersive regime we are considering. 
Using an improved expression would not affect the expression of the scattering coefficients 
and would barely alter the expression of the modes in $x$-space. 

In $x$-space, $\phi_\om^{\rm in}$ is given by the inverse Fourier transform
\be
\phi_\om^{\rm in}(x) = \int_{\mathbb R} e^{i  kx } \tilde \phi_\om^{\rm in}(k) \frac{dk}{\sqrt{2\pi}}, \label{inverseFT} 
\ee
where the integral is taken along the real $k$ axis.
When evaluating it using a saddle point approximation (something which is equivalent to work with the standard WKB approximation), the  
saddle points exactly correspond to the roots of the dispersion relation \eqref{HJ} (when ignoring the 
co-propagating root $k_\om^{\rm co.}$),
see Fig.\ref{Contour_fig}. 
However, this approximation is valid only for $\om \gg \kappa$, which is not the regime we are interested in.
To go beyond the saddle point approximation, we 
treat the exponential $\exp(i \om/\kappa \ln k)$ as a ``slowly varying amplitude'' instead of a ``rapidly oscillating phase''. 
As explained in Sec.III.C of~\cite{Coutant11}, this drastically extends 
the validity range of the result so as to include the limit $\om \ll \kappa$. 

To evaluate the integral in this case, we need to deform the real line of integration of \eq{inverseFT} 
 into the contour in the complex $k$-plane which is depicted on Fig.~\ref{Contour_fig}, left panel. 
Along the ${\mathcal C}_1$ and ${\mathcal C}_2$ parts, 
a saddle point approximation is valid because the locations of the two saddles are $k=\mp k_{\rm saddle}$ where $k_{\rm saddle} = (6\kappa x/c_0 d_0^2)^{1/2}$ is
a high wave number irrespectively of the value of $\om/\kappa \ll 1$. 
These two saddle points correspond to the zero frequency limit 
of the short wavelength roots $k_\om^{\rm out}(x)< 0$ and $-k_{-\om}^{\rm out}(x)>0 $ 
evaluated in the near the blocking point.  
The contributions of these saddles describe respectively the two outgoing reflected waves 
$\varphi_\om^{\rm out}(x)$ and $(\varphi_{-\om}^{\rm out}(x))^*$. 
On the other hand, the contribution of the branch cut, which 
describes the incoming long wavelength mode $\varphi_\om^{\rm in}(x)$, 
cannot be correctly evaluated by a saddle point approximation, precisely because it has a very long wavelength 
$k^{\rm in}_\om(x) = \om/\kappa x$ in the limit $\om \to 0$, fixed $x >0$. 
Collecting the three contributions, one gets 
\be
\phi_\om^{\rm in}(x) = \underbrace{\alpha_\om \varphi_\om^{\rm out}(x) + \beta_\om (\varphi_{-\om}^{\rm out}(x))^*}_{\text{saddle point contributions}} \quad + \underbrace{\varphi_\om^{\rm in}(x)}_{\text{branch cut contribution}}. 
\label{resultB}
\ee
In this expression, the functions $\varphi$ designate the normalized waves which asymptote to the modes
of \eq{plw} when $c$ and $v$ become constant. 
The identification of the above coefficients with those entering Eq.~\eqref{Smat} is unambiguous 
because we work with normalized modes, and sufficiently far from the blocking point. 
(This point is discussed below.) 

It is now easy to verify that the the coefficients $\alpha_\om$ and $\beta_\om$,
which weigh respectively the positive and negative energy outgoing waves, 
agree up to the flip of the sign of $k$ from negative to positive value. 
Taking into account the branch cut of $\ln(k)$ in \eq{inpmode}, 
their ratio is given by
\be
\left|\frac{\beta_\om}{\alpha_\om}\right| = \left| e^{i\frac\om\kappa \ln(e^{i \pi})}\right| = e^{-\frac{\pi \om}{\kappa}}.
\label{HRrat}
\ee
This gives \eq{HRbeta}. For the interested reader, we signal that this derivation is closely 
related to the original works~\cite{Hawking75,Unruh76}, as explained in~\cite{Brout95}.

To complete our analysis,
we give the expressions of the three waves entering \eq{resultB}
in the region where the linearized expression $c - v_x \sim \kappa x$ holds.
Using the standard WKB approximation, their expressions are given by 
\bsub \bea
\varphi_\om^{\rm out}(x) &\sim& \frac{e^{-i \frac23 ( x/d_{\rm br})^{3/2}} e^{-i\frac{\omega}{2\kappa} \ln(x/d_{\rm br})}}{\sqrt{8\pi \kappa (x/d_{\rm br})^{3/2}}} \\
(\varphi_{-\om}^{\rm out}(x))^* &\sim& \frac{e^{i \frac23 ( x/d_{\rm br})^{3/2}} e^{-i\frac{\omega}{2\kappa} \ln(x/d_{\rm br})}}{\sqrt{8\pi \kappa (x/d_{\rm br})^{3/2}}} \\
\varphi_\om^{\rm in}(x) &\sim& \frac{|x/d_{\rm br}|^{i\frac{\om}{\kappa}}}{\sqrt{4\pi \om}} . 
\eea
\label{expli}
\esub
Since, they only depend on $x/d_{\rm br}$, 
this demonstrates that, for low $\om/\kappa$, $d_{\rm br}$ of \eq{broad} is the only 
characteristic length which governs the modes near the blocking point. 

It is also important to remember that while the expressions of \eq{expli} 
make use of the WKB approximation (in position space), 
the scattering coefficients $ \beta_\om$ and $\alpha_\om$
do not because their values follow from the evaluation of the integral along the cut. 
In addition, it should be noticed that the corrections to \eq{expli} 
decrease when the size of the region (where $c - v \sim \kappa x$ holds) 
 increases. 
This size is controlled by the parameter $D$ of Fig.\ref{vprofile_fig}. 
This means that the residual errors induced by 
identification of the coefficients of \eq{resultB} with the
asymptotic scattering coefficients of \eq{Smat} also decrease 
when $D$ increases. 
For more details on this, we refer to Sec.III.C of ~\cite{Coutant11}.

To conclude, we use the above results to derive \eq{Airy}. 
Considering the limit $\om \to 0$ in \eq{inverseFT}, 
one obtains a primitive integral of the Airy function. Then, when acting on it with the derivative of \eq{deltah} to compute the corresponding $\delta h(x)$, we get the Airy function  
of \eq{Airy}.

\newpage
\bibliographystyle{utphys}
\bibliography{Biblio}

\end{document}